# RF power generation

*R.G. Carter*
Engineering Department, Lancaster University, Lancaster LA1 4YR, U.K.
and The Cockcroft Institute of Accelerator Science and Technology, Daresbury, UK

**Abstract**
This paper reviews the main types of r.f. power amplifiers which are, or may be, used for particle accelerators. It covers solid-state devices, tetrodes, inductive output tubes, klystrons, magnetrons, and gyrotrons with power outputs greater than 10 kW c.w. or 100 kW pulsed at frequencies from 50 MHz to 30 GHz. Factors affecting the satisfactory operation of amplifiers include cooling, matching and protection circuits are discussed. The paper concludes with a summary of the state of the art for the different technologies.

## 1 Introduction

All particle accelerators with energies greater than 20 MeV require high-power radio-frequency (r.f.) sources Ref. [1]. These sources must normally be amplifiers in order to achieve sufficient frequency and phase stability. The frequencies employed range from about 50 MHz to 30 GHz or higher. Power requirements range from 10 kW to 2 MW or more for continuous sources and up to 150 MW for pulsed sources. Figure 1 shows the main features of a generic r.f. power system. The function of the power amplifier is to convert d.c. input power into r.f. output power whose amplitude and phase is determined by the low-level r.f. input power. The r.f. amplifier extracts power from high-charge, low-energy electron bunches. The transmission components (couplers, windows, circulators etc.) convey the r.f. power from the source to the accelerator, and the accelerating structures use the r.f. power to accelerate low-charge bunches to high energies. Thus the complete r.f. system can be seen as an energy transformer which takes energy from high-charge, low-energy electron bunches and conveys it to low-charge, high-energy bunches of charged particles. When sufficient power cannot be obtained from a single amplifier then the output from several amplifiers may be combined. In some cases power is supplied to a number of accelerating cavities from one amplifier.

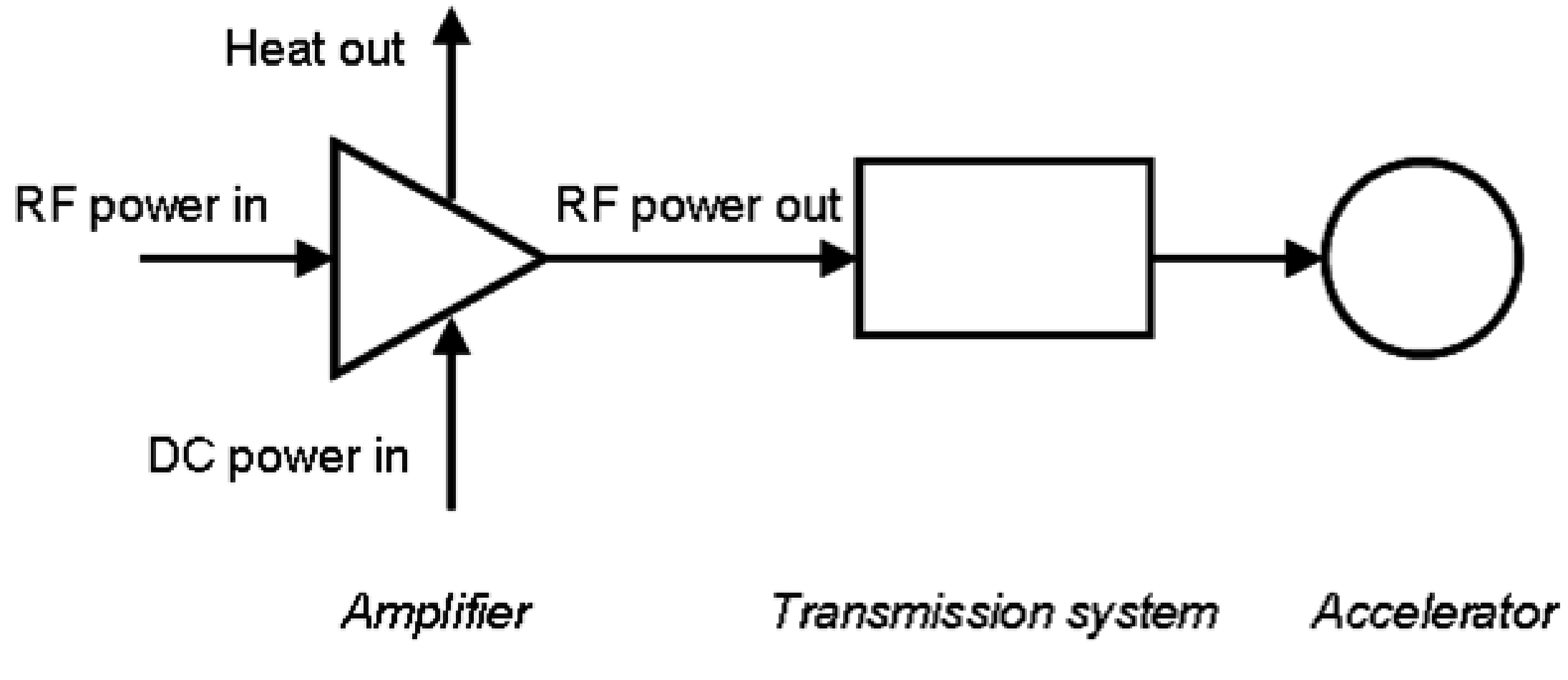

**Fig.1:** Block diagram of the high power r.f. system of an accelerator

Because the amplifier is never completely efficient there is always some conversion of energy into heat. The principle of conservation of energy requires that, in the steady state, the total input and output power must balance, that is

$$P_{RF\,in} + P_{DC\,in} = P_{RF\,out} + P_{Heat} \,. \tag{1}$$

Strictly speaking, the input power should also include power required for other purposes related to the amplifier including heaters, magnets, and cooling systems. However, in many cases these are small compared with the r.f. output power and may be neglected to a first approximation. The efficiency $(\eta_e)$ of the amplifier is the ratio of the r.f. output power to the total input power

$$\eta_e = \frac{P_{RF\,out}}{P_{DC\,in} + P_{RF\,in}} \approx \frac{P_{RF\,out}}{P_{DC\,in}} \,. \tag{2}$$

In many cases the r.f. input power is small compared with the d.c. input power so that it may be neglected to give the approximation shown in Eq. (2). The efficiency is usually expressed as a percentage. The heat which must be dissipated is

$$P_{Heat} = (1 - \eta_e) P_{DC} \,. \tag{3}$$

The other main parameter of the generic amplifier is its gain in decibels given by

$$G = 10 \log_{10} \left( \frac{P_{RF\,out}}{P_{RF\,in}} \right). \tag{4}$$

The physics of the energy exchange between the electron bunches and the r.f. power means that the size of the space in which the exchange takes place must be small compared with the distance an electron moves in one r.f. cycle. Thus the size of an amplifier decreases with decreasing d.c. voltage and with increasing frequency. However, for a given r.f. output power and efficiency, the energy density within the amplifier increases with decreasing size. The need to keep the working temperature of the amplifier below the level at which it will cease to operate reliably means that the maximum possible output power from a single device is determined by the working voltage, the frequency, and the technology employed. Other factors which are important in the specification of power amplifiers include reliability and, in some cases, bandwidth.

The capital and running cost of an accelerator is strongly affected by the r.f. power amplifiers in a number of ways. The capital cost of the amplifiers (including replacement tubes) is an appreciable part of the total capital cost of the accelerator. Their efficiency determines the electricity required and, therefore, the running cost. The gain of the final power amplifier determines the number of stages required in the r.f. amplifier chain. The size and weight of the amplifiers determines the space required and can, therefore, have an influence on the size and cost of the tunnel in which the accelerator is installed.

All r.f. power amplifiers employ either vacuum-tube or solid-state technology. Vacuum tubes use d.c. or pulsed voltages from several kV to hundreds of MV depending upon the type of tube, the power level, and the frequency. The electron velocities can be comparable with the velocity of light and the critical tube dimensions are therefore comparable with the free-space wavelength at the working frequency. Vacuum tubes can therefore generate r.f. power outputs up to 1 MW continuous wave (c.w.) and 150 MW pulsed. The main tube types employed in accelerators are: gridded tubes (tetrodes); inductive output tubes (IOTs); klystrons and magnetrons. The magnetron is an oscillator rather than an amplifier and its use is currently restricted to medical linacs. The gyrotron, which is widely used as a power source in fusion reactors, has the potential to be used for high-frequency accelerators. Solid-state r.f. power transistors operate at voltages from tens to hundreds of volts. The electron mobility is much less in semiconductor materials than in vacuum and the device sizes are therefore small and the power which can be generated by a single transistor is of the order of hundreds of watts continuous and up to 1 kW pulsed. Large numbers of transistors must be operated in parallel to reach even the lowest power levels required for accelerators. This paper reviews the state of the art of these types of amplifier and discusses some of the factors affecting their successful operation.

## 2 Solid-state amplifiers

Solid-state power amplifiers for use at high frequencies employ transistors fabricated from wide band-gap semiconductor materials (Si, GaAs, GaN, SiC, diamond) Refs. [2,3]. Table 1 shows the parameters of some commercially available state-of-the-art r.f. power transistors.

**Table 1:** State-of-the-art r.f. power transistors

| | **LR 301**[a] | **TGF2023-20** | | **TGF4240-SCC** |
|---|---|---|---|---|
| Manufacturer | Polyfet | TriQuint | | TriQuint |
| Material | Si LDMOS | GaN | | GaAs |
| Frequency | 350 MHz | 3.0 GHz | 14 GHz | 8.5 GHz |
| Mean r.f. power | 300 W | 90 W | 50 W | 14 W |
| Operating voltage | 28 V | 30 V | 30 V | 8 V |
| Gain | 13 dB | 17 dB | 5 dB | 10 dB |
| Maximum junction temperature | 200 °C | 200 °C | 200 °C | 150 °C |

a. Two transistors in push-pull.

The state of the art of solid-state r.f. power amplifiers for use in accelerators is represented by the 352 MHz amplifier built for the SOLEIL light source [4]. This amplifier combines the outputs from four 50 kW towers to generate a total output power of 180 kW. The towers are constructed from 726, 315 W modules each of which includes two Si LDMOS transistors operated in push-pull. In this mode the transistors conduct alternately for half the r.f. cycle so that each is operated in class B (see Section 3.2) with a theoretical efficiency of 78%. Figure 2 shows the arrangement of the 172 modules in one tower. The gain of one module is 13 dB and the complete amplifier has 53 dB gain and an overall efficiency of the order of 50%.

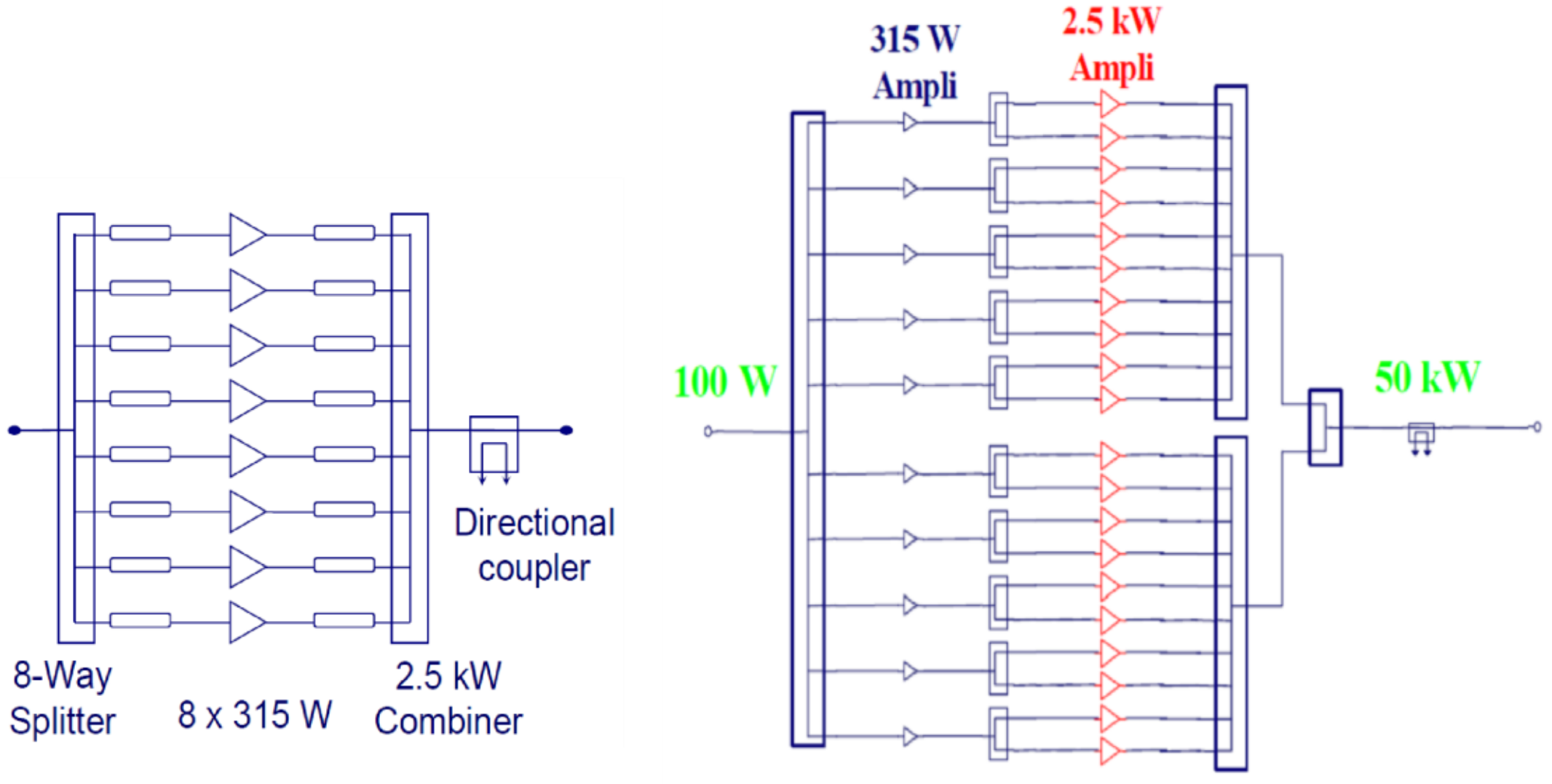

(a) (b)

**Fig. 2:** Arrangement of the SOLEIL 352 MHz amplifier, (a) 2.5 kW amplifier and, (b) 50 kW tower (© IEEE 2005 reproduced with permission)

It is sometimes claimed that solid-state amplifiers are more reliable than their vacuum-tube counterparts by a factor of as much as 2.5 [5]. Because many transistors are operated in parallel, the failure of one produces a negligible drop in power output. However, the transistors are operated close to their design limits and they are vulnerable to accidental overloads. Appreciable power is dissipated in the combining circuits and provision may have to be made to isolate a device which fails so that it does not adversely affect the performance of others nearby. Other advantages claimed for solid-state amplifiers are their high stability, low maintenance requirements, absence of warm-up time, and low-voltage operation. The supply voltage is low enough to avoid the high-voltage and X-ray safety issues arising with tube amplifiers. The penalty which must be paid for this is the need to supply and handle very large d.c currents using large bus-bars in which there are appreciable ohmic losses.

## 3 Tetrode amplifiers

Tetrode vacuum tubes are well established as high-power r.f. sources in the VHF (30–300 MHz) band. The arrangement of a 150 kW, 30 MHz tetrode is shown in Fig. 3 (from Ref. [6]). The construction is coaxial with the cathode inside and the anode outside. The output power available from such a tube is limited by the maximum current density available from the cathode and by the maximum power density which can be dissipated by the anode. The length of the anode must be much less than the free-space wavelength of the signal to be amplified in order to avoid variations in the signal level along it. The perimeter of the anode must likewise be much less than the free-space wavelength in order to avoid the excitation of azimuthal higher-order modes in the space between the anode and the screen grid. The spacings between the electrodes must be small enough for the transit time of an electron from the cathode to the anode to be much less than the r.f. period. If attempts are made to reduce the transit time by raising the anode voltage then there may be flashover between the electrodes. The bulk of the heat which must be dissipated arises from the residual kinetic energy of the electrons as they strike the anode. Thus provision must be made for air or liquid cooling of the anode (See section 9.1). Note the substantial copper anode with channels for liquid cooling shown in Fig. 3. For further information on gridded tubes see Ref. [7].

The current of electrons emitted from the cathode surface is controlled by the field of the control grid modified by those of the other two electrodes. The dependence of the anode current on the electrode voltages is given approximately by

$$I_a \approx C\left(V_{g1} + \frac{V_{g2}}{\mu_2} + \frac{V_a}{\mu_a}\right)^n \tag{5}$$

where $V_{g1}, V_{g2},$ and $V_a$ are, respectively, the potentials of the control grid, the screen grid, and the anode with respect to the cathode and $C$, $\mu_1, \mu_2,$ and $n$ are constants [8]. Typically $\mu_2 \sim 5 - 10$, $\mu_a \sim 100 - 200$ and $n$ is in the range 1.5 to 2.5.

Figure 4 shows the characteristic curves of a typical tetrode [9]. The control grid voltage is plotted against the anode voltage, both being referred to the cathode. The three sets of curves show the anode current (solid lines), control grid current (dashed lines), and the screen-grid current (chain dotted lines). The voltages of the anode (known as the plate in the USA) and the screen grid are positive with respect to the cathode. The curves shown are for a fixed screen-grid voltage of +900 V. It is clear that the anode current depends strongly on the control grid voltage and more weakly on the anode voltage. The control grid voltage is normally negative with respect to the cathode to prevent electrons being collected on the grid with consequent problems of heat dissipation. The screen grid, which is maintained at r.f. ground, prevents capacitive feedback from the anode to the control grid. The screen grid voltage is typically about 10% of the d.c. anode voltage. If the anode voltage falls below that of the screen grid then any secondary electrons liberated from the anode are collected by the screen grid. Thus when tetrodes are operated as power amplifiers the anode voltage is always greater than the screen grid voltage.

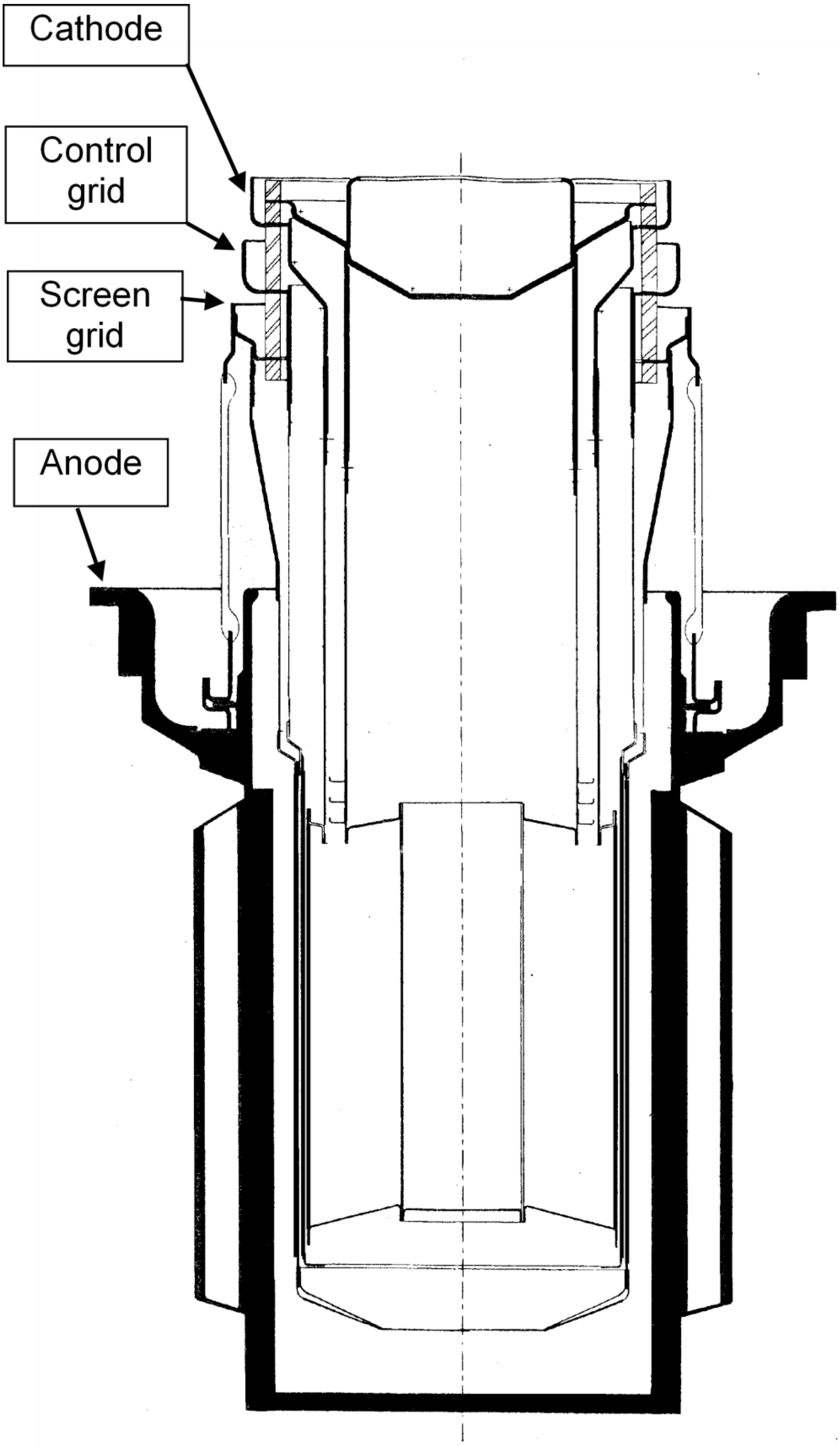

**Fig. 3:** Cross-sectional view of a high-power tetrode (Courtesy of e2v technologies)

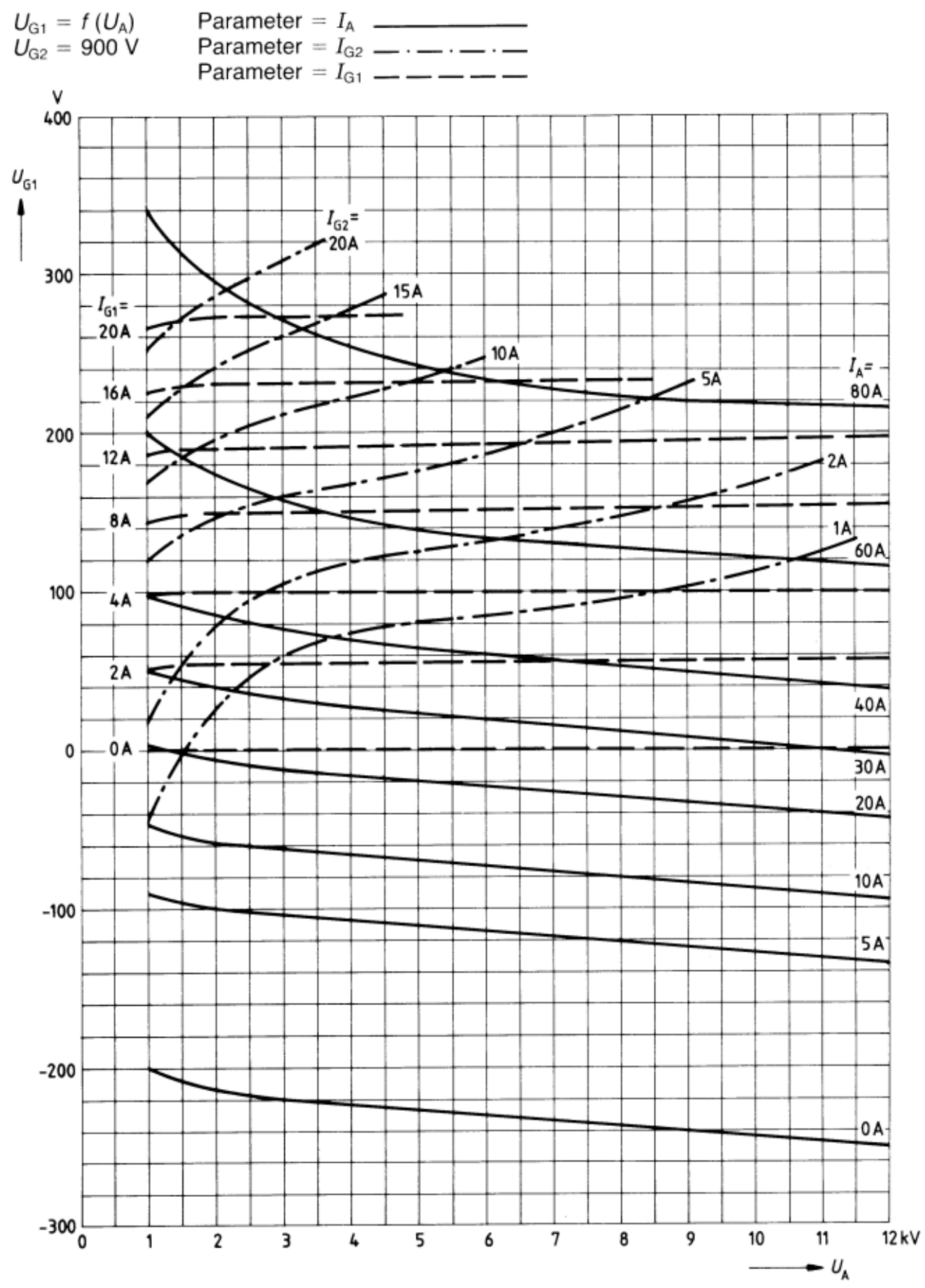

**Fig. 4:** Characteristic curves of the RS 2058 CJ tetrode for $V_{g2}$ = 900 V (Courtesy of Siemens AG)

### 3.1 Tetrode amplifier circuits

Figure 5 shows the circuit of a grounded-grid tetrode amplifier with a tuned anode circuit. At low frequencies a resistive anode load may be used but this is unsatisfactory in the VHF band and above because of the effects of parasitic capacitance. The amplifiers used in accelerators are operated at a single frequency at any one time so the limited bandwidth of the tuned anode circuit is not a problem. At the resonant frequency the load in the anode circuit comprises the shunt resistance of the resonator ($R_S$) in parallel with the load resistance ($R_L$). If the load impedance has a reactive component then it merely detunes the resonator and can easily be compensated for. The d.c. electrode potentials are maintained by the power supplies shown and the capacitors provide d.c. blocking and r.f. bypass.

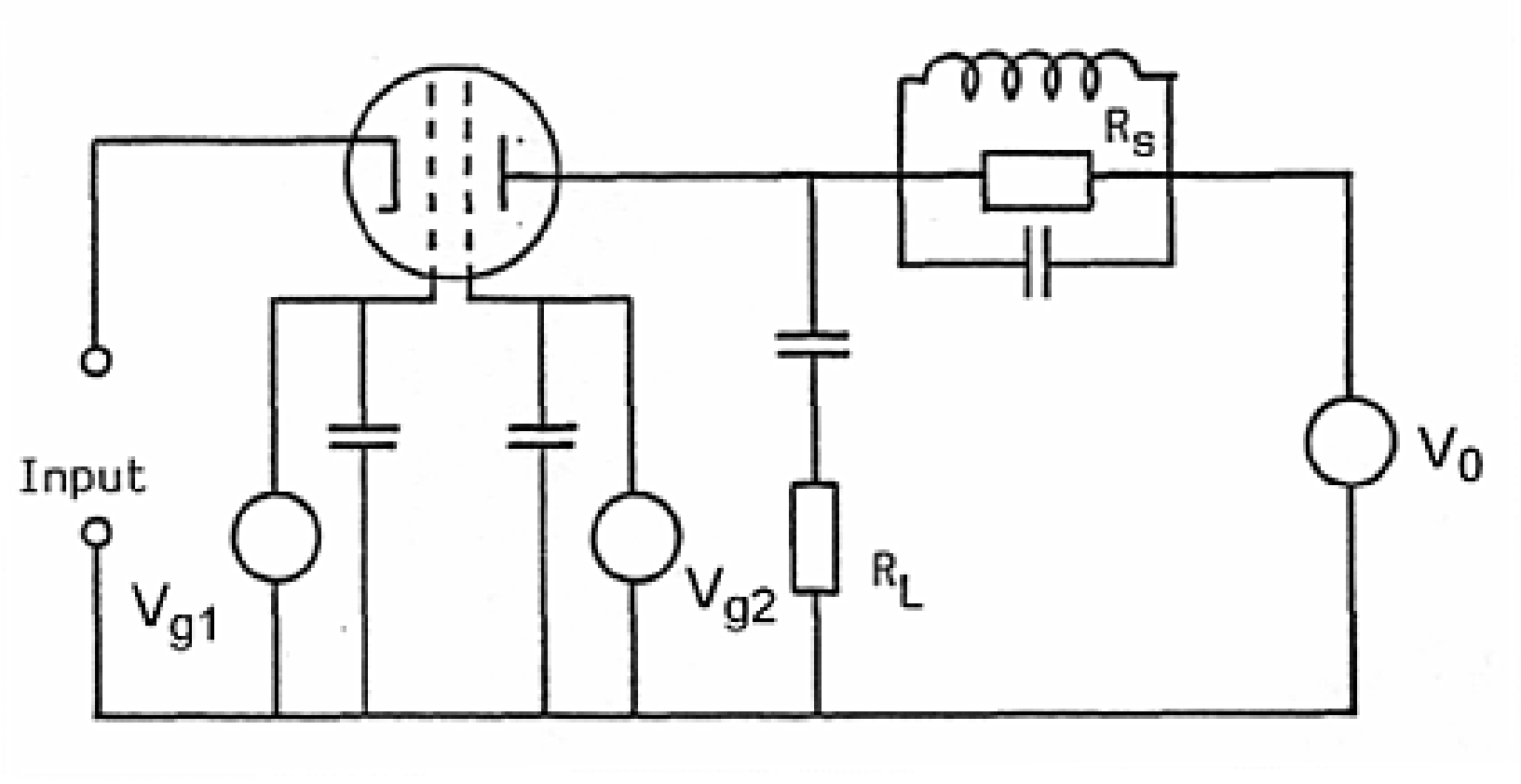

**Fig. 5:** Circuit of a grounded-grid tetrode amplifier

### 3.2 Class B operation

The operation of tuned amplifiers is designated as class A, B, or C according to the fraction of the r.f. cycle for which the tube is conducting. Amplifiers for accelerators are normally operated in, or close to, class B in which the tube conducts for half of each cycle. To achieve this the d.c. bias on the control grid is set so that the anode current is just zero in the absence of an r.f. input signal. This ensures that the tube conducts only during the positive half-cycle of the r.f. input voltage and the conduction angle is 180 degrees. Figure 6 illustrates the current and voltage waveforms, normalized to their d.c. values, under the simplifying assumptions that the behaviour of the tube is linear and that the anode load is tuned to resonance.

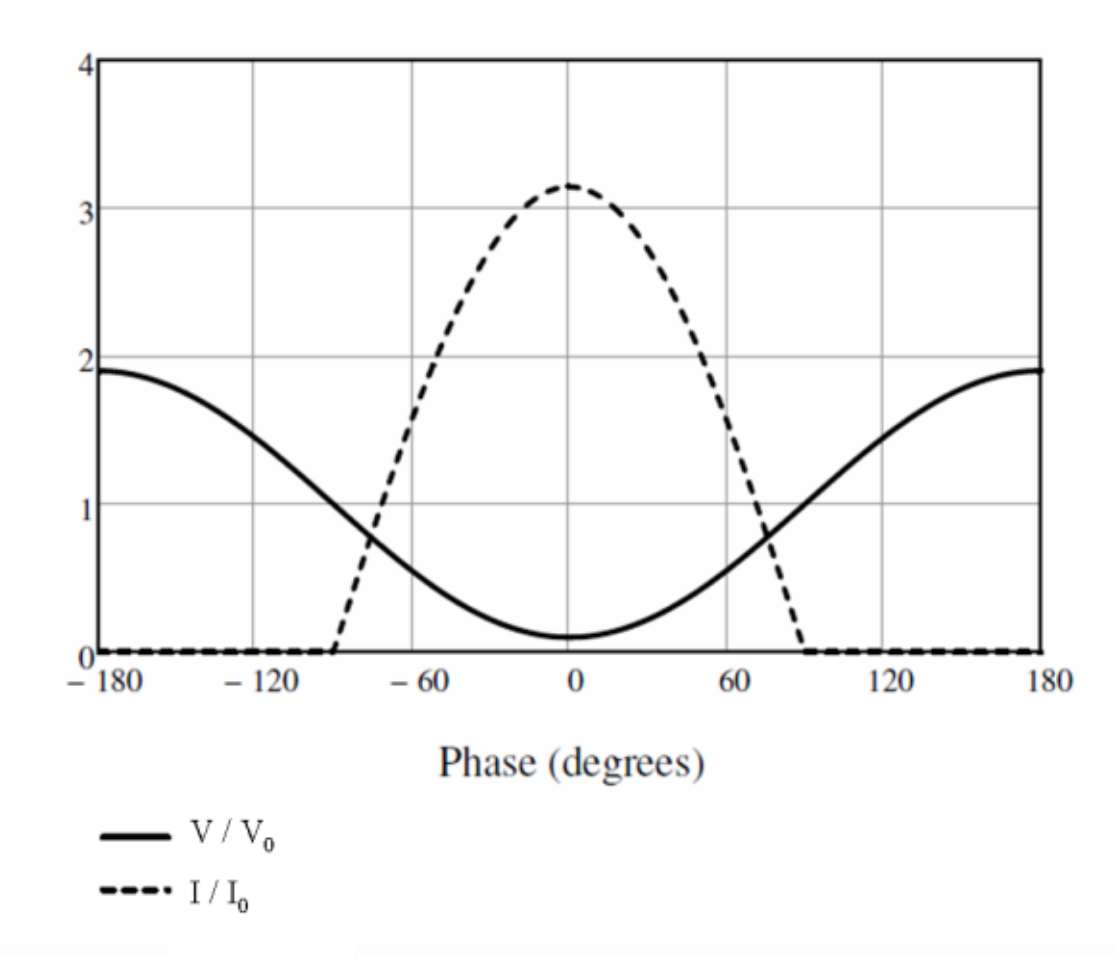

**Fig.6:** Current and voltage waveforms in a tetrode amplifier operating in class B

If it is assumed that the tube is linear whilst it is conducting, the d.c. anode current is found by Fourier analysis of the current waveform in Fig. 6 to be

$$I_0 = \frac{I_{pk}}{\pi} = 0.318 \tag{6}$$

where $I_{pk}$ is the peak current during the cycle. Similarly the r.f. anode current is given by

$$I_2 = 0.5\, I_{pk} \ . \tag{7}$$

In an ideal class B amplifier the minimum r.f. voltage is zero so that the relationship between the r.f. and d.c. anode voltages is

$$V_2 = V_0 \ . \tag{8}$$

The d.c power is given by

$$P_0 = V_0 I_0 = \frac{V_0 I_{pk}}{\pi} \tag{9}$$

and the r.f. output power by

$$P_2 = \frac{1}{2} V_2 I_2 = \frac{1}{4} V_0 I_{pk} \ . \tag{10}$$

The theoretical efficiency is then

$$\eta_e = \frac{P_2}{P_0} = \frac{\pi}{4} = 78.5\% \ . \tag{11}$$

In practice the efficiency of a tetrode amplifier is less than the theoretical limit for two reasons. First, the non-linear relationship between the sinusoidal control-grid voltage and the anode current shown by Eq. (5) means that the constants in Eqs. (6) and (7) become 0.278 and 0.458 when $n = 1.5$ and 0.229 and 0.397 when $n = 2.5$ Second, the need to ensure that the anode voltage always exceeds the screen grid voltage means that the amplitude of the r.f. anode voltage $V_2$ is limited to around 90% of $V_0$. Substitution of these revised figures into Eqs. (9)–(11) shows that the practical efficiency may be expected to lie in the range 74% to 78% depending on the value of *n*.

### 3.3 Class A, AB, and C operation

The proportion of the r.f. cycle during which the tetrode is conducting can be adjusted by changing the d.c. bias voltage on the control grid. The operation then falls into one of a number of classes as shown in Table 2. If the negative grid bias is reduced then anode current flows when there is no r.f. drive. The application of a small r.f. drive voltage produces class A amplification in which the tube conducts throughout the r.f. cycle. As the r.f. drive voltage is increased the tube becomes cut off for part of the cycle and the operation is intermediate between class A and class B and known as class AB. When the grid bias is made more negative than that required for class B operation the tube conducts for less than half the r.f. cycle and the operation is described as class C. Analysis similar to that given above shows that the ratio of the r.f. anode current to the d.c. anode current increases as the conduction angle is reduced and, therefore, the efficiency increases. However, the amplitude of the r.f. drive voltage required to produce a given amplitude of the r.f. anode current increases as the conduction angle is reduced so that the gain of the amplifier decreases. Finally, the harmonic content of the r.f. anode current waveform increases as the conduction angle increases. The properties of the different classes of amplifier are summarized in Table 2. The amplifiers used for particle accelerators should ideally have high efficiency, high gain and low harmonic output. For this reason it is usual to operate them in class B or in class AB with a conduction angle close to 180°. It should be noted that solid-state amplifiers can also be operated in the classes shown in Table 2.

**Table 2:** Classes of amplifier

| Class | Conduction angle | Maximum theoretical efficiency | Negative grid bias increasing | Gain increasing | Harmonics increasing |
|---|---|---|---|---|---|
| A | 360° | 50% | ↓ | ↑ | ↓ |
| AB | 180° – 360° | 50% – 78% | | | |
| B | 180° | 78% | | | |
| **C** | < 180° | 78% – 100% | | | |

### 3.4 Tetrode amplifier design

The process by which a tetrode amplifier can be designed is best explained by means of an example. This is based upon a 62 kW, 200 MHz amplifier used in the CERN SPS [10]. The example was chosen because sufficient information is available about the amplifier to verify the results of the calculations. The amplifier uses a single RS2058CJ tetrode [9]) operating with a d.c. anode voltage of 10 kV and 900 V screen grid bias. The design procedure described below is based upon that given in Ref. [11].

The actual amplifier is operated in class AB but quite close to class B. For simplicity class B operation is assumed in the calculations which follow. The first stage is to estimate the probable efficiency of the amplifier. We will assume that the minimum anode voltage is 1.5 kV. Scaling the figures given above which take account of the non-linearity of the tetrode suggests that the efficiency of the amplifier will lie in the range 70% to 74%. Let us assume that the efficiency is 72%. This figure can be adjusted later, if necessary, when the actual efficiency has been calculated. Then the d.c. power input necessary to obtain the desired output power is

$$P_0 = 62/0.72 = 86\,\text{kW}\,. \tag{12}$$

The d.c. anode voltage was chosen to be 10 kV so the mean anode current is

$$I_0 = 86/10 = 8.6\,\text{A}\,. \tag{13}$$

The theoretical value of $I_{pk}$ is given by Eq. (6) but when the non-linearity of the tetrode is taken into account it is found that this figure lies in the range 3.6 to 4.4 depending upon the value of *n*. If we take the factor to be 4.0 then

$$I_{pk} = 4.0 \times 8.6 = 34\,\text{A}\,. \tag{14}$$

Next we construct the load line on the characteristic curves for the tube shown in Fig. 7 by joining the point 1.5 kV, 34 A to the quiescent point (10 kV, 0 A). We note that this requires the control grid voltage to swing slightly positive with a maximum of +70 V.

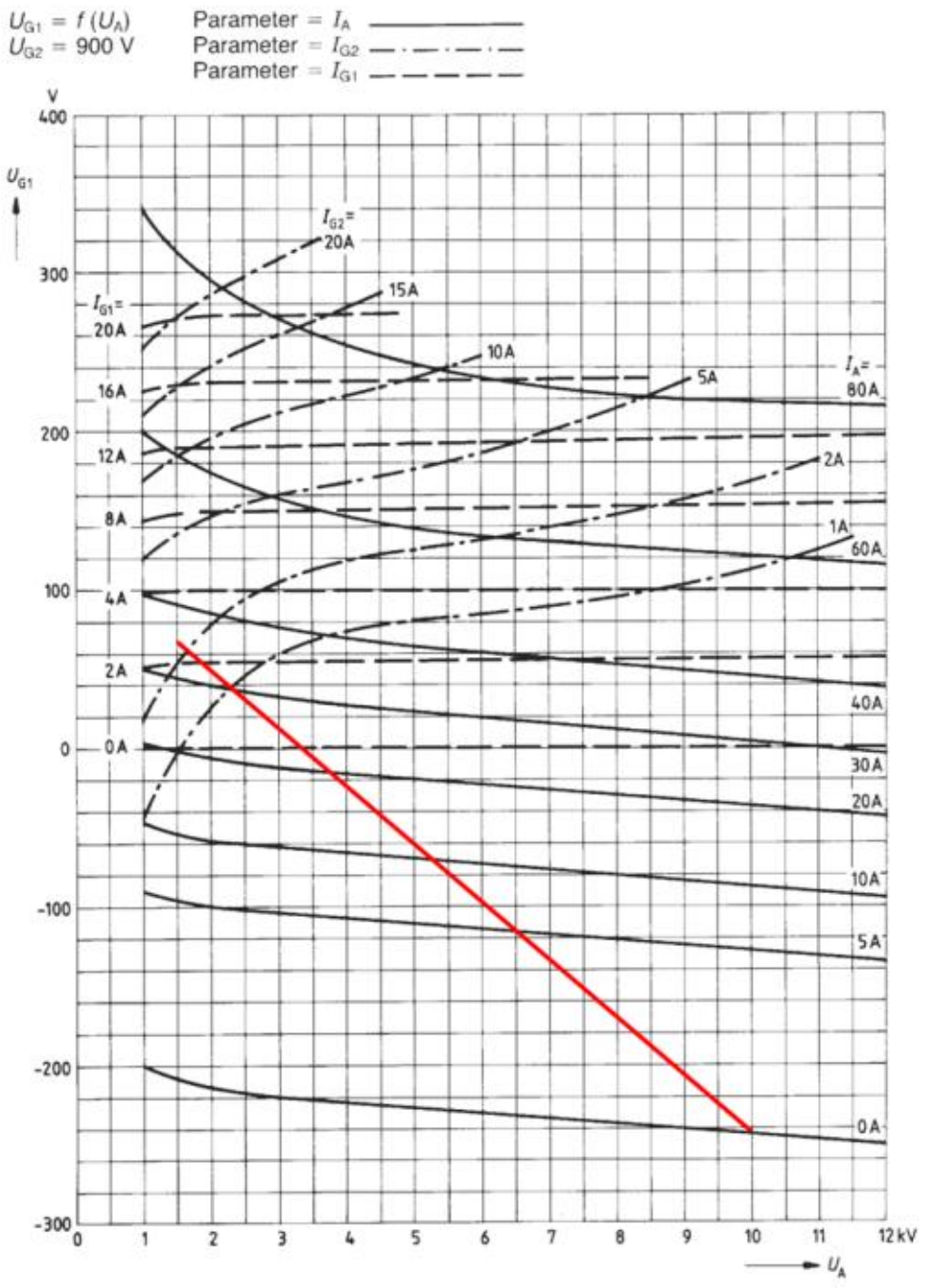

**Fig. 7:** Tetrode characteristic curves with load line (Courtesy of Siemens AG)

The d.c. and r.f. currents are found by numerical Fourier analysis of the anode current waveform using values read from Fig. 7 at intervals of 15°. The anode voltages are given by

$$V_a = 10 - 8.5\cos\theta \tag{15}$$

where $\theta$ is the phase angle and the results are shown in Table 3.

**Table 3:** Anode currents at 15° phase intervals taken from Fig. 7

| **Phase (degrees)** | $V_a$ **(kV)** | $I_a$ **(A)** |
|---|---|---|
| 0 | 1.5 | 34 |
| 15 | 1.8 | 32.5 |
| 30 | 2.6 | 28 |
| 45 | 4.0 | 18 |
| 60 | 5.7 | 8 |
| 75 | 7.8 | 3 |
| 90 | 10.0 | 0 |

The d.c. and r.f. anode currents can be found by Fourier analysis of the current waveform using the numerical formulae given in Ref. [11]:

$$I_0 = \frac{1}{12}(0.5I_{a0} + I_{a15} + I_{a30} + I_{a45} + I_{a60} + I_{a75}) \tag{16}$$

and

$$I_2 = \frac{1}{12}(I_{a0} + 1.93I_{a15} + 1.73I_{a30} + 1.41I_{a45} + I_{a60} + 0.52I_{a75}) \tag{17}$$

where the subscripts refer to the phase angles in degrees. When these formulae are used with the data from Table 3 the results are

$$I_0 = 8.9 \text{ A} \tag{18}$$

and

$$I_2 = 15.0 \text{ A}. \tag{19}$$

Thus the d.c. input power is

$$P_0 = I_0 V_0 = 89 \text{ kW}. \tag{20}$$

The amplitude of the r.f. voltage is

$$V_2 = 10.0 - 1.5 = 8.5 \text{ kV} \tag{21}$$

and the r.f. output power is

$$P_2 = \frac{1}{2} V_2 I_2 = 64 \text{ kW} \tag{22}$$

which is very close to the desired value and gives an efficiency of 72% as originally assumed. The effective load resistance is

$$R_L = V_2 / I_2 = 570\,\Omega. \tag{23}$$

The source impedance of the output of the amplifier can be found by noting that if the r.f. load resistance is zero the anode voltage is constant and the peak anode current is 46 A for the same r.f. voltage on the control grid. Thus the short circuit r.f. current is 20 A and the anode source resistance $(R_a)$ is 1.7 kΩ.

To find the input impedance of the amplifier we note that the amplitude of the r.f. control grid voltage is

$$V_1 = 245 + 70 = 315 \text{ V} \tag{24}$$

and that for grounded grid operation the amplitude of the r.f. input current is

$$I_1 = I_2 + I_{g1RF} \approx I_2. \tag{25}$$

The amplitude of the r.f. control grid current $(I_{g1RF})$ may be obtained by reading the control grid currents off Fig. 7 at 15° intervals and employing Eq. (17). The result is 0.67 A which is small compared with the r.f. anode current and can be neglected in the first approxmation. The r.f. input resistance is

$$R_1 = V_1 / I_1 = 20\,\Omega. \tag{26}$$

Finally we note that the input power is

$$P_1 = \frac{1}{2} V_1 I_1 = 2.5 \text{ kW} \tag{27}$$

and that the power gain of the amplifier is

$$\text{Gain} = 10 \log (64/2.5) = 14 \text{ dB} . \quad (28)$$

Table 4 shows a comparison between the figures calculated above and those reported in Ref. [10]. The differences between the two columns of Table 4 are attributable to the difference between the actual class AB operation and the class B operation assumed in the calculations.

**Table 4:** Comparison between actual and calculated parameters of the amplifier described in Ref. [10]

| Parameter | Actual | Calculated | |
|---|---|---|---|
| $V_0$ | 10 | 10 | kV |
| $I_0$ | 9.4 | 8.9 | A |
| $V_{g2}$ | 900 | 900 | V |
| $V_{g1}$ | −200 | −245 | V |
| $P_{out}$ | 62 | 64 | kW |
| $P_{in}$ | 1.8 | 2.5 | kW |
| Gain | 15.4 | 14 | dB |
| $\eta$ | 64 | 72 | % |

### 3.5 Practical details

Figure 8 shows a simplified diagram of a tetrode amplifier. The tube is operated in the grounded grid configuration with coaxial input and output circuits. The outer conductors of the coaxial lines are at ground potential and they are separated from the grids by d.c. blocking capacitors. The anode resonator is a re-entrant coaxial cavity which is separated from the anode by a d.c. blocking capacitor. The output power is coupled through an impedance matching device to a coaxial line. The anode HT connection and cooling water pipes are brought in through the centre of the resonator.

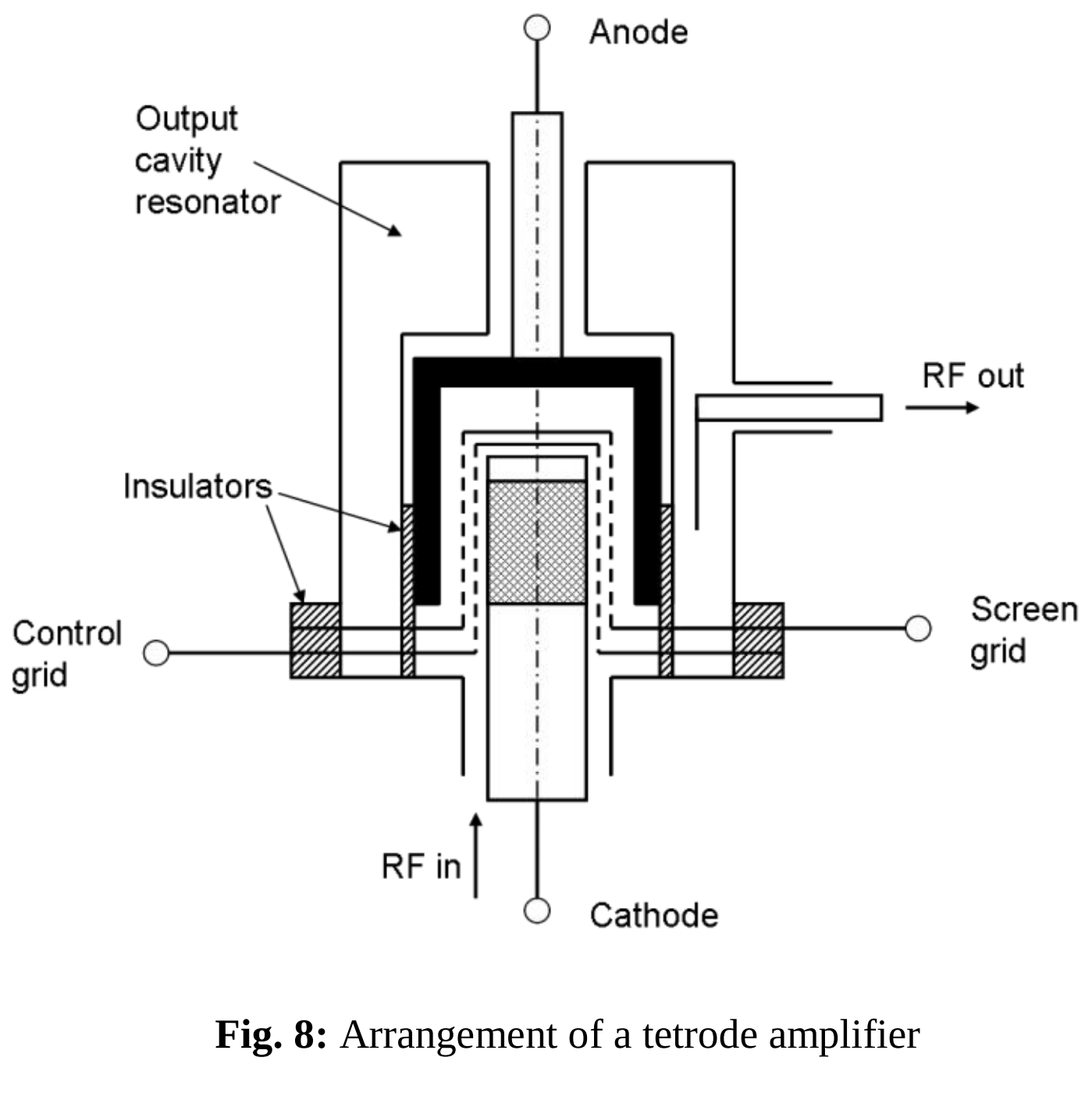

**Fig. 8:** Arrangement of a tetrode amplifier

The electrodes of the tube form coaxial lines with characteristic impedances of a few ohms. We have seen above that the input impedance of the amplifier is typically a few tens of ohms and the output impedance a few hundred ohms. Thus the terminations of both the input and output lines are close to open-circuits. The anode resonator therefore has one end open-circuited and the other short-circuited and it must be an odd number of quarter wavelengths long at resonance. Typically the resonator is 3/4 of a wavelength long. In that case the point at which the output coaxial line is coupled into the resonator can be used to transform the impedance to provide a match. There is also a voltage node towards the lower end of the outer part of the resonator and this can be used to bring connections through to the anode [12]. The higher-order modes of the cavity can be troublesome and it is usually necessary to damp them by the selective placing of lossy material or of coupling loops connected to external loads within the cavity [10, 12–14]. The tube heater connections must incorporate some means of decoupling from the r.f. circuit [10, 13].

It will be clear from what has already been said that the tube input and output are mismatched to the external connections. It is therefore necessary to devise matching networks for these connections. Figure 9(a) shows a simplified form of the input circuit of the amplifier discussed in Section 3.3. The source (normally 50 ohms) feeds the tube through a length of low-impedance coaxial line. The line is terminated by the tube input impedance shunted by the capacitance between the control grid and the cathode. For the RS2058CJ tetrode this capacitance is 140 pF so that it has a reactance of 5.7 ohms at 200 MHz. This is comparable with the input resistance of 20 ohms so its effect must be allowed for. In practice the connecting coaxial line is made up of several sections having different impedances [12, 13].

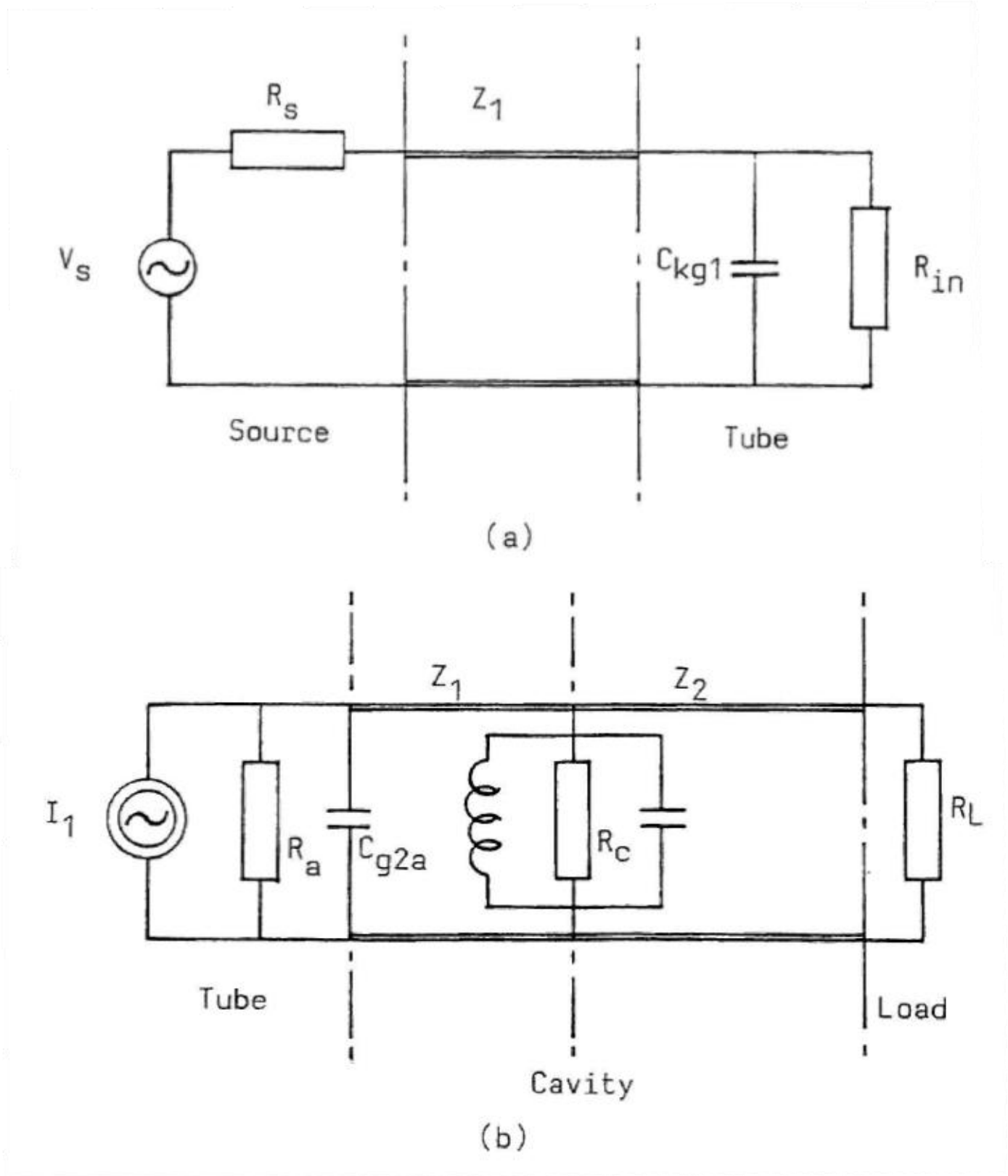

**Fig. 9:** (a) Input circuit and, (b) output circuit of a tetrode amplifer

Figure 9(b) shows the output circuit of the amplifier. The output resistance is shunted by the capacitance between the screen grid and the anode which is 40 pF, with a reactance of 20 ohms at 200 MHz. The tube is connected to the coaxial cavity resonator by a coaxial line and the load is connected by another line and, possibly, an impedance-matching network. The effect of the anode/screen grid capacitance is normally to tune the frequency of the cavity somewhat. Variations in the load impedance may have a similar effect which can be compensated for by tuning the cavity. They may also require a variable matching network so that the load can be kept correctly matched to the amplifier [9, 12, 13]. If the impedance presented to the anode is too high the anode voltage swing may be excessive with the possibility of damage to the tube through internal arcing.

### 3.6 Operation of tetrode amplifiers in parallel

When higher powers are required than can be obtained from a single tube then it is possible to operate several tubes in parallel. Two such systems are described in Ref. [14]. The original four 500 kW, 200 MHz power amplifiers for the CERN SPS each comprised four 125 kW tetrodes operating in parallel. Figure 10 shows the arrangement of one amplifier. The loads on the fourth arms of the 3 dB couplers normally receive no power. If one tube fails, however, they must be capable of absorbing the power from the unbalanced coupler. The amplifier also contains coaxial transfer switches (not shown in Fig. 10) which make it possible for a faulty tube to be completely removed from service. The remaining three tubes can then still deliver 310 kW to the load. A more recent design of 500 kW amplifier for the same accelerator employs sixteen 35 kW units operated in parallel with a seventeenth unit as the driver stage. Both types of amplifier operate at anode efficiencies greater than 55% and overall efficiencies greater than 45%.

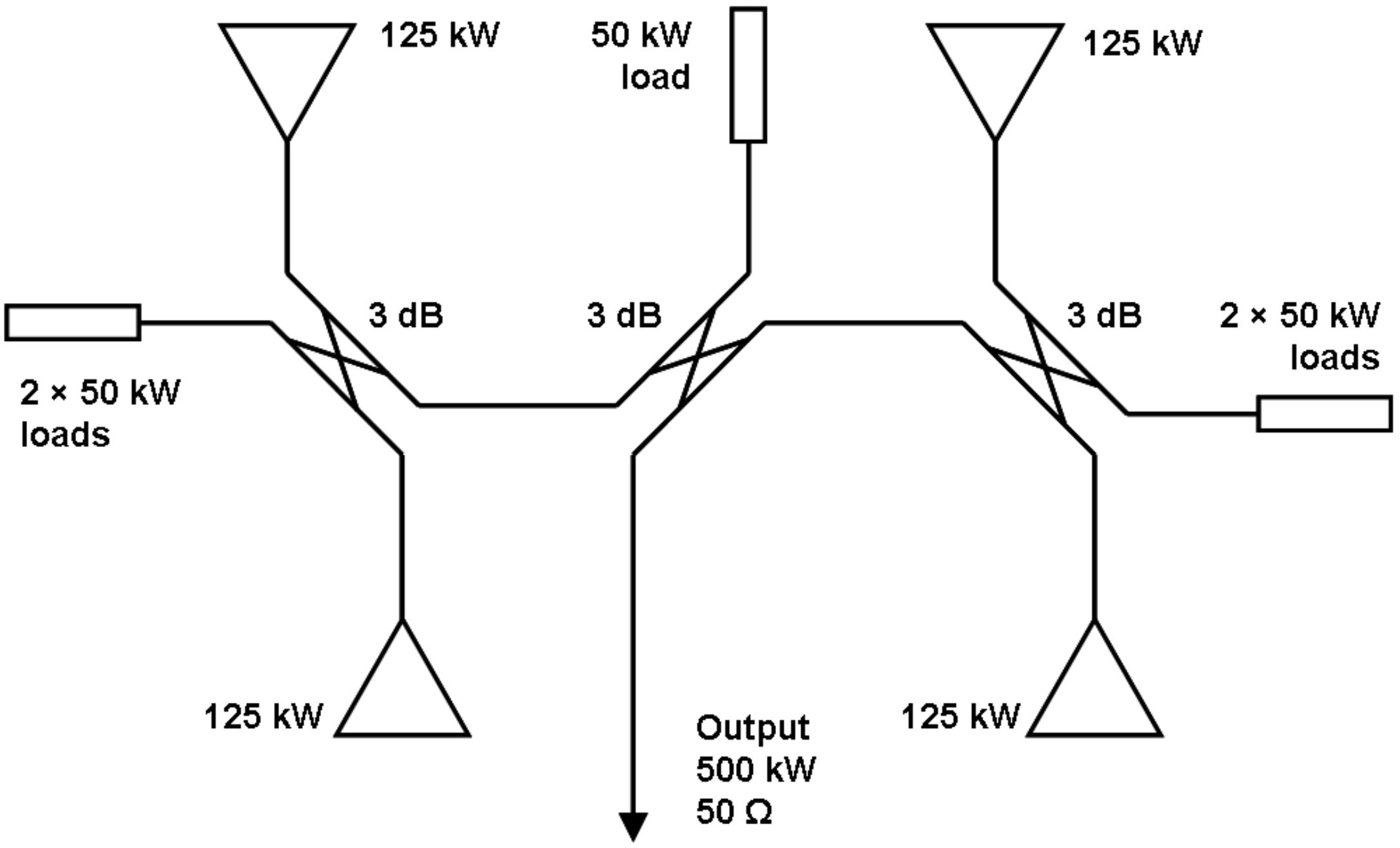

**Fig. 10:** Arrangement for operation of tetrode amplifiers in parallel (Courtesy of Siemens AG)

### 3.7 The Diacrode®

A recent development of the tetrode is the Diacrode [15]. In this tube the coaxial line formed by the anode and the screen grid is extended to a short circuit as shown in Fig.11. The consequence of this change is that the standing wave now has a voltage anti-node, and a current node, at the centre of the active region of the tube. The tetrode, in contrast, has a voltage anti-node and current node just beyond the end of the active region, as shown in Fig.11. Thus, for the same r.f. voltage difference between the anode and the screen grid, the Diacrode has a smaller reactive current flow and much smaller power dissipation in the screen grid than a tetrode of similar dimensions. This means that, compared with conventional tetrodes, Diacrodes can either double the output power at a given operating frequency or double the frequency for a given power output. The gain and efficiency of the Diacrode are the same as those of a conventional tetrode. Table 5 shows the comparison between the TH 526 tetrode and the TH 628 Diacrode operated at 200 MHz.

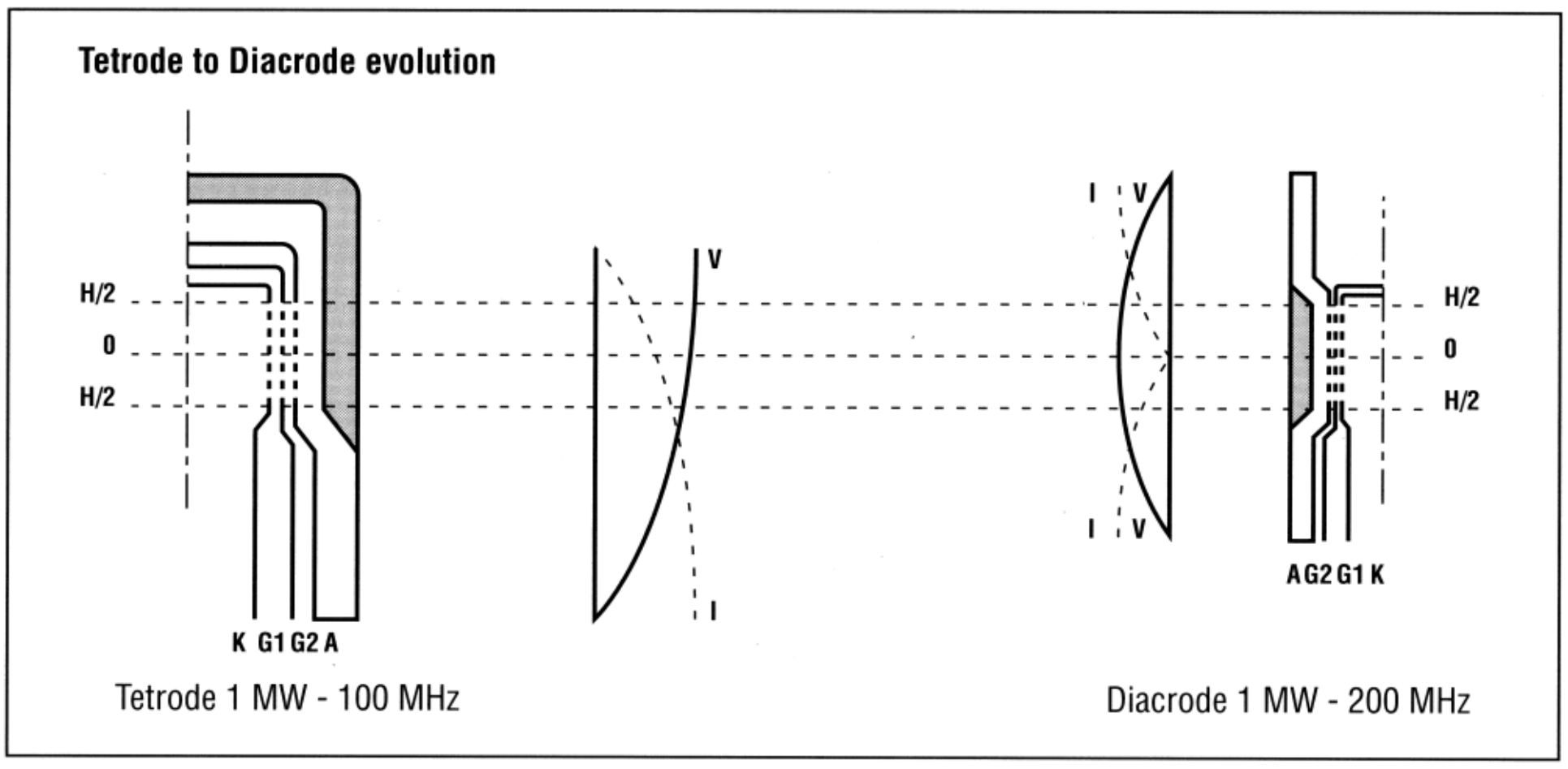

**Fig.11:** Comparison between a tetrode and a Diacrode® (Courtesy of Thales Electron Devices)

**Table 5:** Comparison between the TH 526 tetrode and the TH 628 Diacrode® at 200 MHz

| | **TH 526** | | **TH 628** | |
|---|---|---|---|---|
| Pulse duration (ms) | 2.2 | c.w. | 2.5 | c.w. |
| Peak output power (kW) | 1600 | – | 3000 | – |
| Mean output power (kW) | 240 | 300 | 600 | 1000 |
| Anode voltage (kV) | 24 | 11.5 | 26 | 16 |
| Anode current (A) | 124 | 75 | 164 | 96 |
| Peak input power (kW) | 64.9 | – | 122.5 | – |
| Mean input power (kW) | – | 21 | – | 32 |
| Gain (dB) | 13.9 | 11.5 | 13.9 | 15 |

## 4 Inductive output tubes

The tetrode suffers from the disadvantage that the same electrode, the anode, is part of both the d.c. and the r.f. circuits. The output power is, therefore, limited by screen grid and anode dissipation. In addition the electron velocity is least when the current is greatest because of the voltage drop across the output resonator (see Fig. 6). To get high power at high frequencies it is necessary to employ high-velocity electrons and to have a large collection area for them. It is therefore desirable to separate the electron collector from the r.f. output circuit. The possibility that these two functions might be separated from each other was originally recognised by Haeff in 1939 but it was not until 1982 that a commercial version of this tube was described [16]. Haeff called his invention the 'Inductive Output Tube' (IOT) but it is also commonly known by the proprietary name Klystrode®.

Figure 12 shows a schematic diagram of an IOT [17]. The electron beam is formed by a gridded, convergent-flow electron gun and confined by an axial magnetic field (not shown). The gun is biased so that no current flows except during the positive half-cycle of the r.f. input. Thus electron bunches are formed and accelerated through the constant potential difference between the cathode and the anode. The bunches pass through a cavity resonator as shown in Fig. 13 so that their azimuthal magnetic field induces a current in the cavity (hence the name of the tube). Because the cavity is tuned to the repetition frequency of the bunches, the r.f. electric field in the interaction gap is maximum in the retarding sense when the centre of a bunch is at the centre of the gap. The interaction between the bunches and the cavity resonator is similar to that in a class B amplifier. Figure 14 shows a plot of the positions of typical electrons against time. The slopes of the lines are proportional to the electron velocities and they show how kinetic energy is extracted from the electron bunches as they pass through the output gap (indicated by dashed lines). The r.f. power transferred to the cavity is equal to the kinetic power given up by the electrons. Because the electron velocity is high it is possible to use a much longer output gap than in a tetrode. The r.f. power passes into and out of the vacuum envelope through ceramic windows.

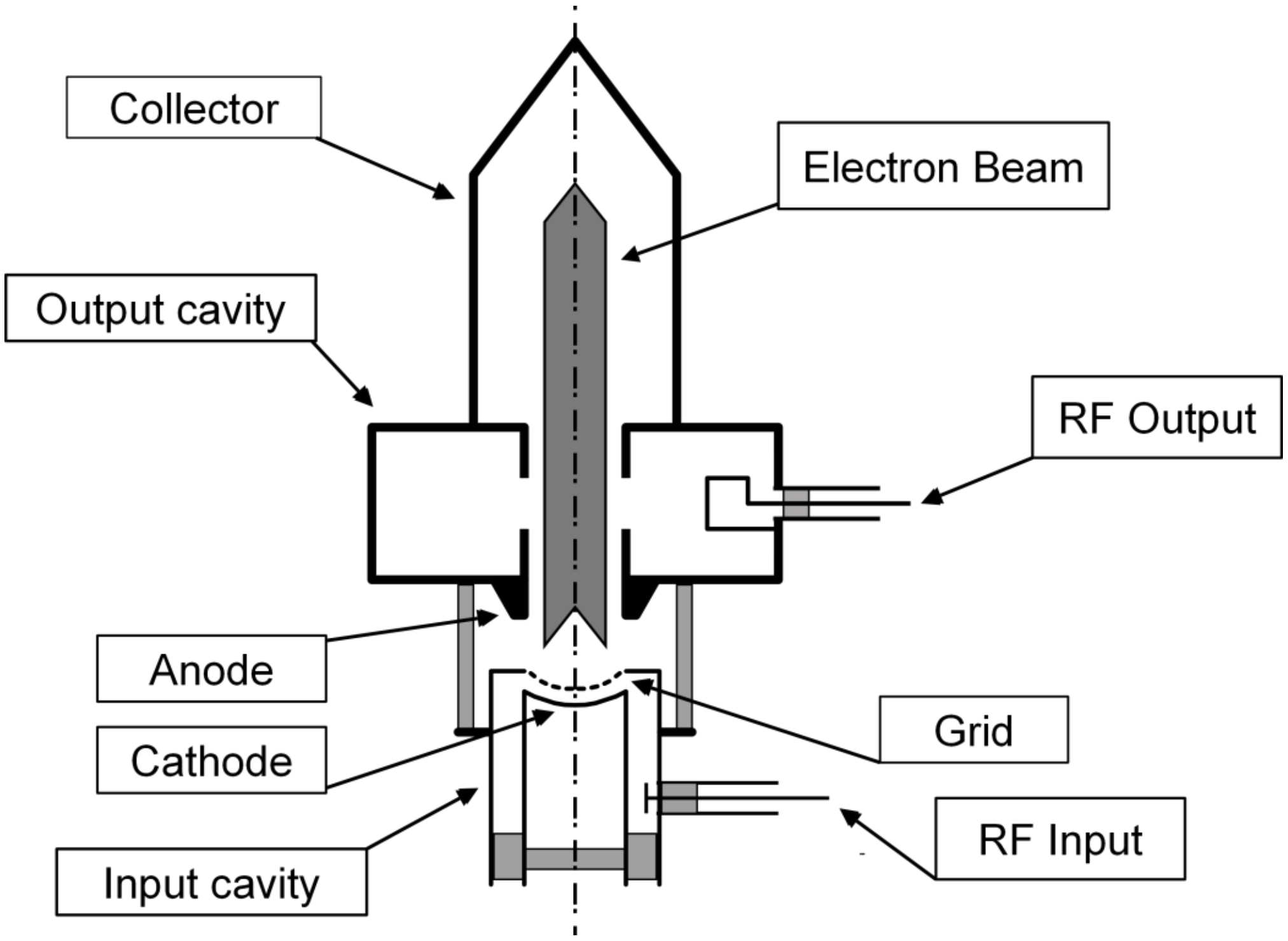

**Fig. 12:** Arrangement of an inductive output tube (IOT) (© IEEE, 2010, reproduced with permission)

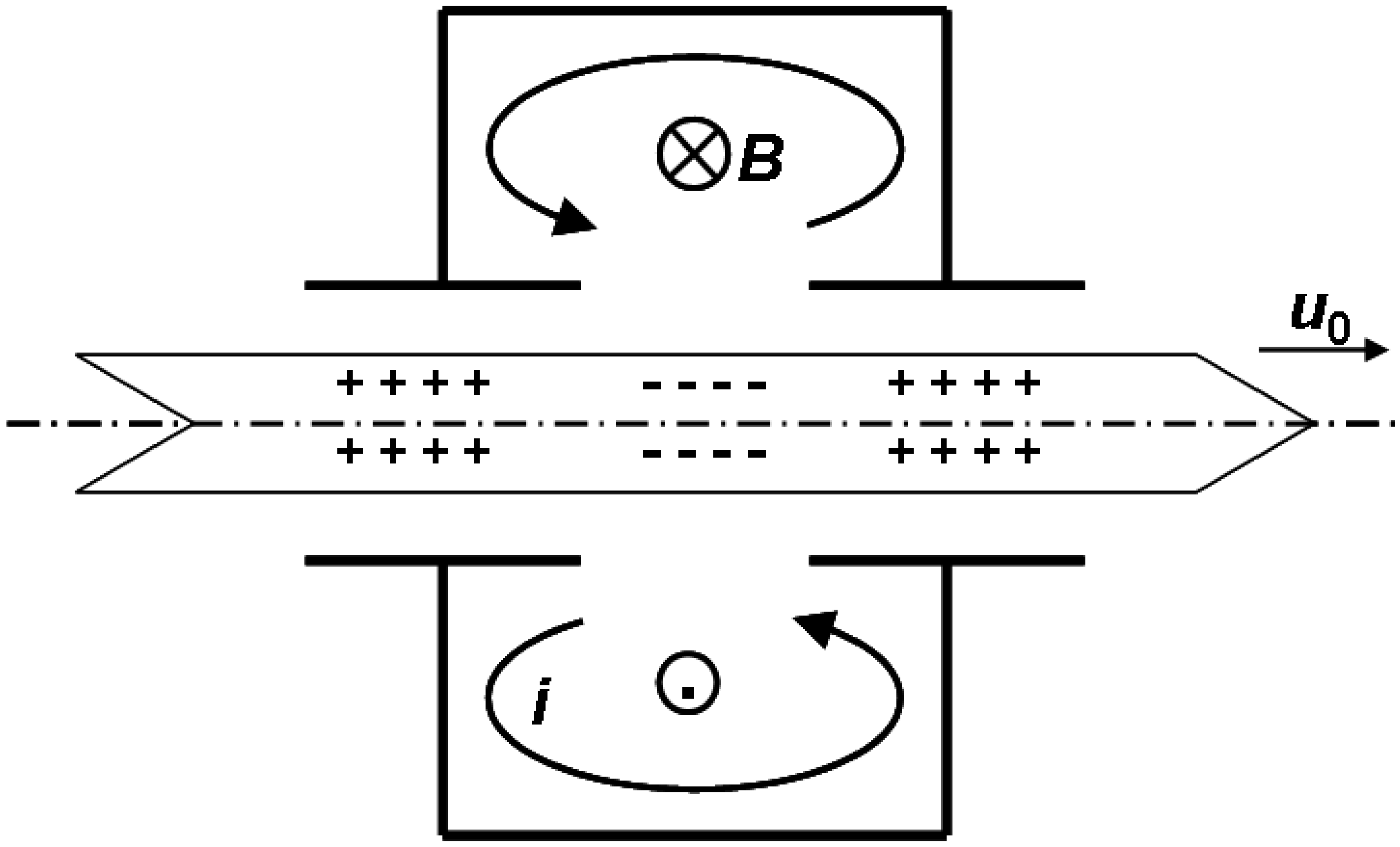

**Fig. 13:** Interaction between a bunched electron beam and a cavity resonator

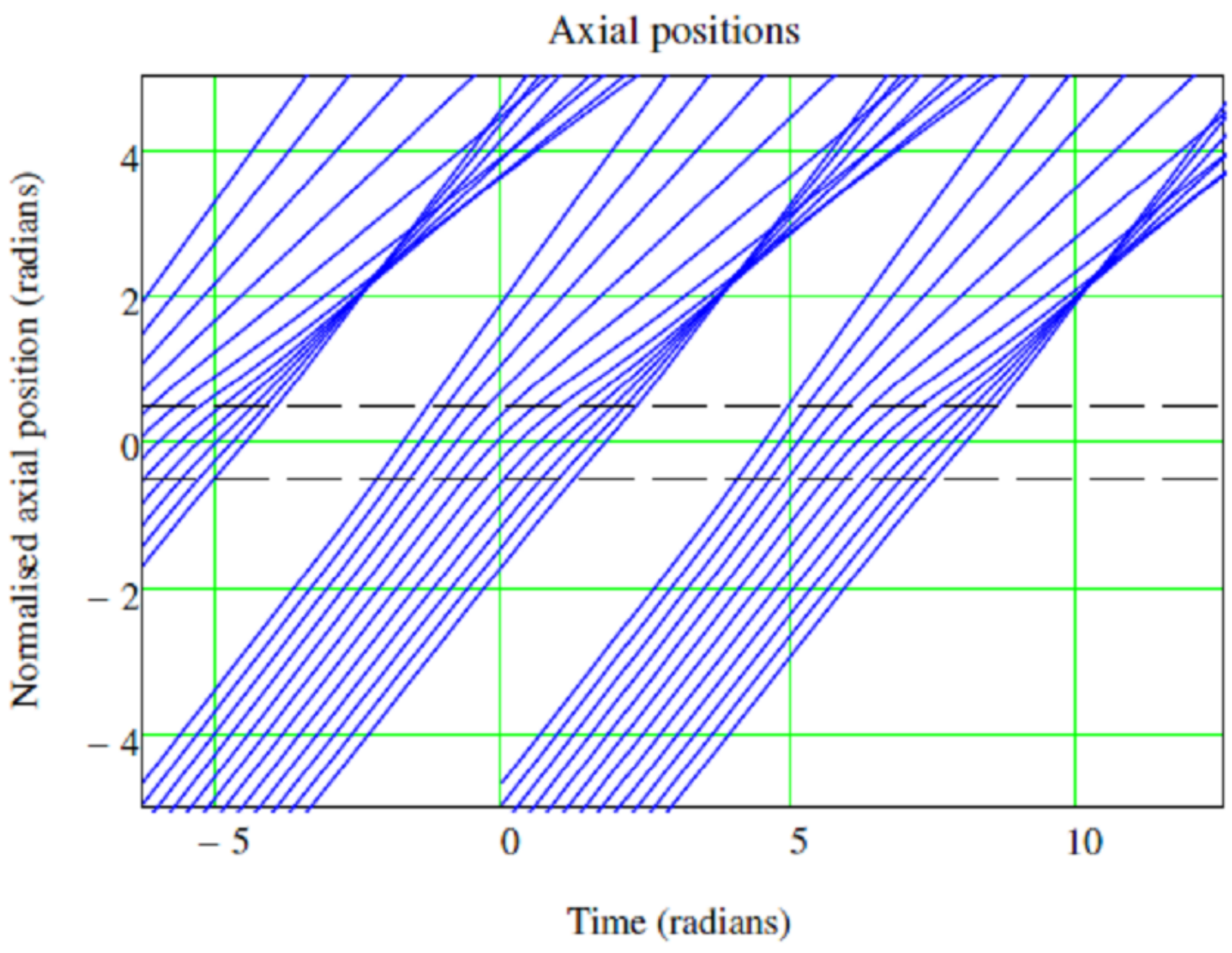

**Fig. 14:** Electron trajectories in an IOT

The efficiency of an IOT can be estimated by noting that the relationship between the r.f. and d.c. currents in the beam is, approximately, from Eqs. (6) and (7)

$$I_1 = \frac{\pi}{2} I_0 \,. \tag{29}$$

The effective voltage of the output gap is the product of the r.f. gap voltage and a transit time factor. The effective gap voltage cannot be greater than about 90% of the voltage used to accelerate the electrons because the electrons leaving the gap must have sufficient residual velocity to enable them to leave the gap and pass into the collector. The maximum r.f. output power is therefore given by

$$P_2 = \frac{1}{2} I_1 V_{gap,eff} = \frac{0.9}{2} \cdot \frac{\pi}{2} \cdot I_0 V_0 = 0.71 P_0 \tag{30}$$

so that the maximum efficiency is approximately 70%.

Advantages of the IOT are that it does not need a d.c. blocking capacitor in the r.f. output circuit because the cavity is at ground potential, and that it has higher isolation between input and output and a longer life than an equivalent tetrode. These advantages are offset to some extent by the need for a magnetic focusing field. The typical gain is greater than 20 dB and is appreciably higher than that of a tetrode; high enough in fact for a 60 kW tube to be fed by a solid-state driver stage. IOTs have been designed for UHF TV applications. The IOTs designed for use in accelerators are operated in class B or class C. In principle the efficiency can be still further enhanced by collector depression. Further information about the IOT can be found in Refs. [16, 18, 19]. Table 6 shows the parameters of some IOTs designed for use in accelerators.

**Table 6:** Parameters of IOTs for use in accelerators

| | **2KDW250PA** | **VKP-9050** | **VKL-9130A** |
|---|---|---|---|
| Manufacturer | CPI/Eimac | CPI | CPI |
| Frequency (MHz) | 267 | 500 | 1300 |
| Beam voltage (kV) | 67 | 40 | 35 |
| Beam current (A) | 6.0 | 3.5 | 1.3 |
| RF output power (kW) | 280 | 90 | 30 |
| Efficiency (%) | 70 | >65 | >65 |
| Gain (dB) | 22 | >22 | >20 |

## 5 Klystrons

At a frequency of 1.3 GHz the continuous output power of an IOT is limited to around 30 kW by the need to use a control grid to modulate the electron beam. At higher frequencies and high powers it is necessary to modulate the beam in some other way. In the klystron this is achieved by passing an un-modulated electron beam through a cavity resonator which is excited by an external r.f. source. The electrons are accelerated or retarded according to the phase at which they cross the resonator and the beam is then said to be velocity modulated. The beam leaving the gap has no current modulation but, downstream from the cavity, the faster electrons catch up the slower ones so that bunches of charge are formed as shown in Fig. 15.

When an output cavity, tuned to the signal frequency, is placed in the region where the beam is bunched the result is the simple two-cavity klystron illustrated in Fig. 16. RF power is induced in the second cavity in exactly the same way as in an IOT. This cavity presents a resistive impedance to the current induced in it by the electron beam so that the phase of the field across the gap is in anti-phase with the r.f. beam current. Electrons which cross the gap within ± 90° of the bunch centre are retarded and give up energy to the field of the cavity. Since more electrons cross the second gap during the retarding phase than the accelerating phase there is a net transfer of energy to the r.f. field of the cavity. Thus the klystron operates as an amplifier by converting some of the d.c. energy input into r.f. energy in the output cavity.

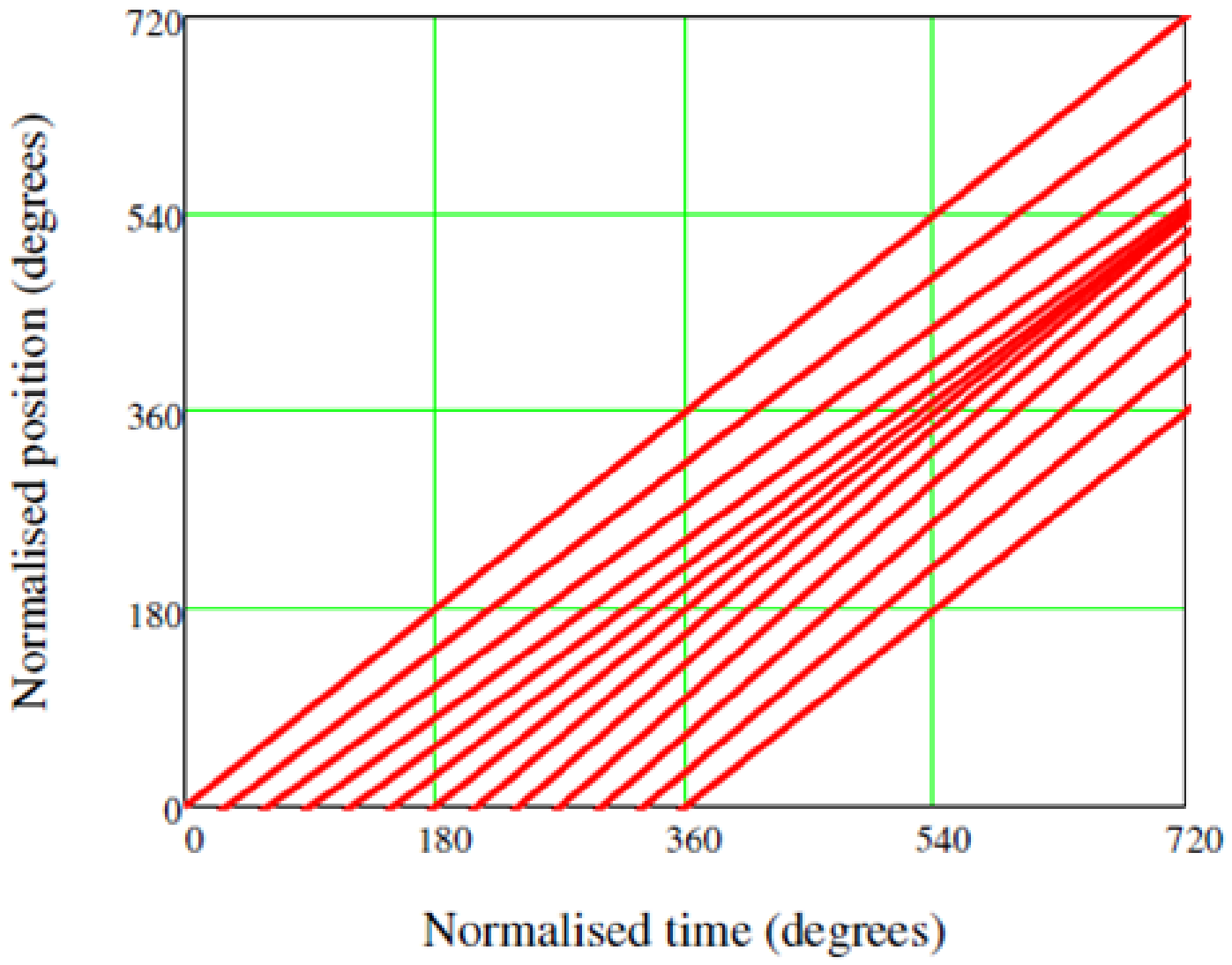

**Fig. 15:** Applegate diagram showing formation of bunches in a velocity-modulated electron beam

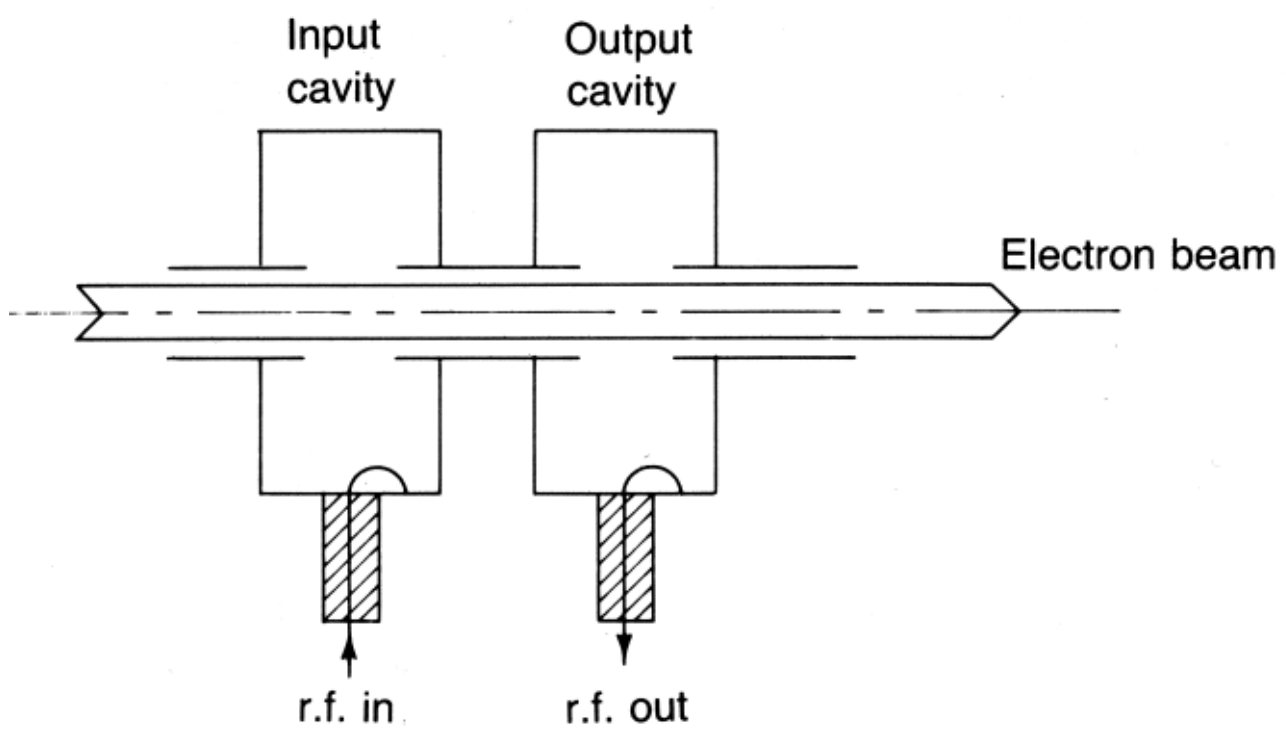

**Fig. 16:** Arrangement of a two-cavity klystron

In practice the gain and efficiency of a two-cavity klystron is too low to be of practical value. It is therefore usual to add further cavity resonators in order to increase the gain, efficiency, and bandwidth of the tube. Figure 17 shows the arrangement of a multi-cavity klystron. The electron beam is formed by a diode electron gun for which

$$I_0 = K V_0^{1.5} \tag{31}$$

where $K$ is a constant known as the perveance which, typically, has a value in the range $0.5 \text{ to } 2.0\times10^{-6}\ \text{AV}^{-1.5}$. The function of all the cavities, except the last, is to form tight electron bunches from which r.f. power can be extracted by the output cavity. The first and last cavities are tuned to the centre frequency and have $Q$ factors which are determined largely by the coupling to the input and output waveguides. The intermediate, or idler, cavities normally have high $Q$ and are tuned to optimize the performance of the tube. The long electron beam is confined by an axial magnetic field to avoid interception of electrons on the walls of the drift tube. The spent electrons are collected by a collector in exactly the same way as in an IOT. The r.f. power passes into and out of the vacuum envelope through ceramic windows.

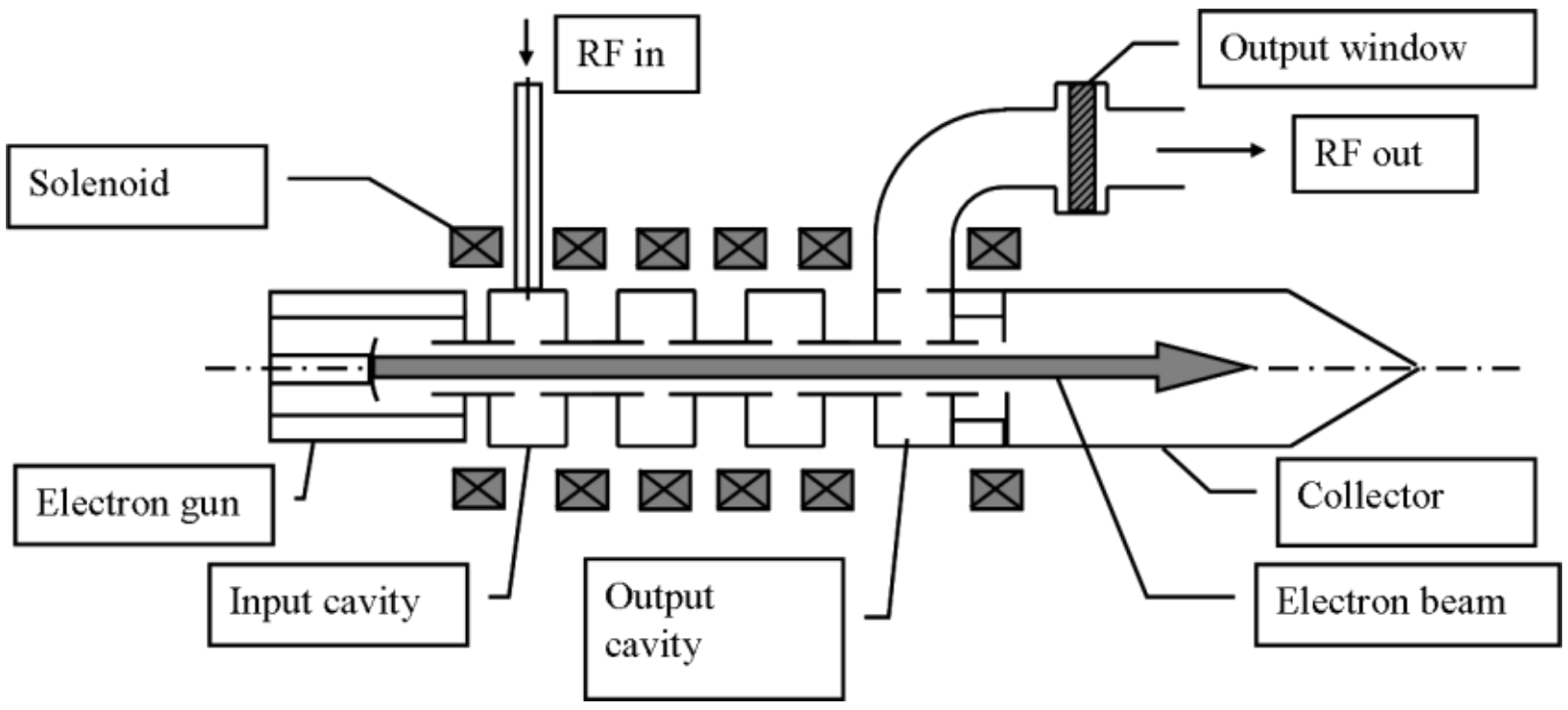

**Fig. 17:** Arrangement of a multicavity klystron

### 5.1 Electron bunching in klystrons

The Applegate diagram in Fig. 15 ignores the effect of space-charge on the bunching. The space-charge forces oppose the bunching and, under small-signal conditions, the beam has current modulation but no velocity modulation at the plane of the bunch. As the beam drifts further, the space-charge forces cause the bunches to disperse and re-form periodically. From the point of view of an observer travelling with the mean electron velocity, the electrons would appear to be executing oscillations about their mean positions at the electron plasma frequency. The plasma frequency is modified to some extent by the boundaries surrounding the beam and by the presence of the magnetic focusing field. The electron plasma frequency is given by

$$\omega_p = \left(\eta\rho/\varepsilon_0\right)^{0.5} \tag{32}$$

where $\eta$ is the charge to mass ratio of the electron and $\rho$ is the charge density in the beam. The distance from the input gap to the first plane at which the bunching is maximum is then a quarter of a plasma wavelength $\left(\lambda_p\right)$ given by

$$\lambda_p = 2\pi\, u_0 \,/\, \omega_p \tag{33}$$

where $u_0$ is the mean electron velocity. Theoretically the second cavity should be placed at a distance $\lambda_p/4$ from the input gap so that the induced current in the second cavity is maximum. In practice it is found that this would make a tube inconveniently long and the distance between the gaps is a compromise between the strength of interaction and the length of the tube.

The bunching length is independent of the input signal except at very high drive levels when it is found that it is reduced. If attempts are made to drive the tube still harder the electron trajectories cross over each other and the bunching is less. Figure 18 shows a typical Applegate diagram for a high power klystron. It should be noted that, in comparison with the diagram in Fig. 15, the axes have been exchanged and uniform motion of the electrons at the initial velocity has been subtracted. The peak accelerating and retarding phases of the fields in the cavities are indicated by + and – signs. Those electrons which cross the input gap at an instant when the field is zero proceed without any change in their velocities and appear as horizontal straight lines. Retarded electrons move upwards and accelerated electrons move downwards in the diagram. Because the cavities are closely spaced space-charge effects are not seen until the final drift region.

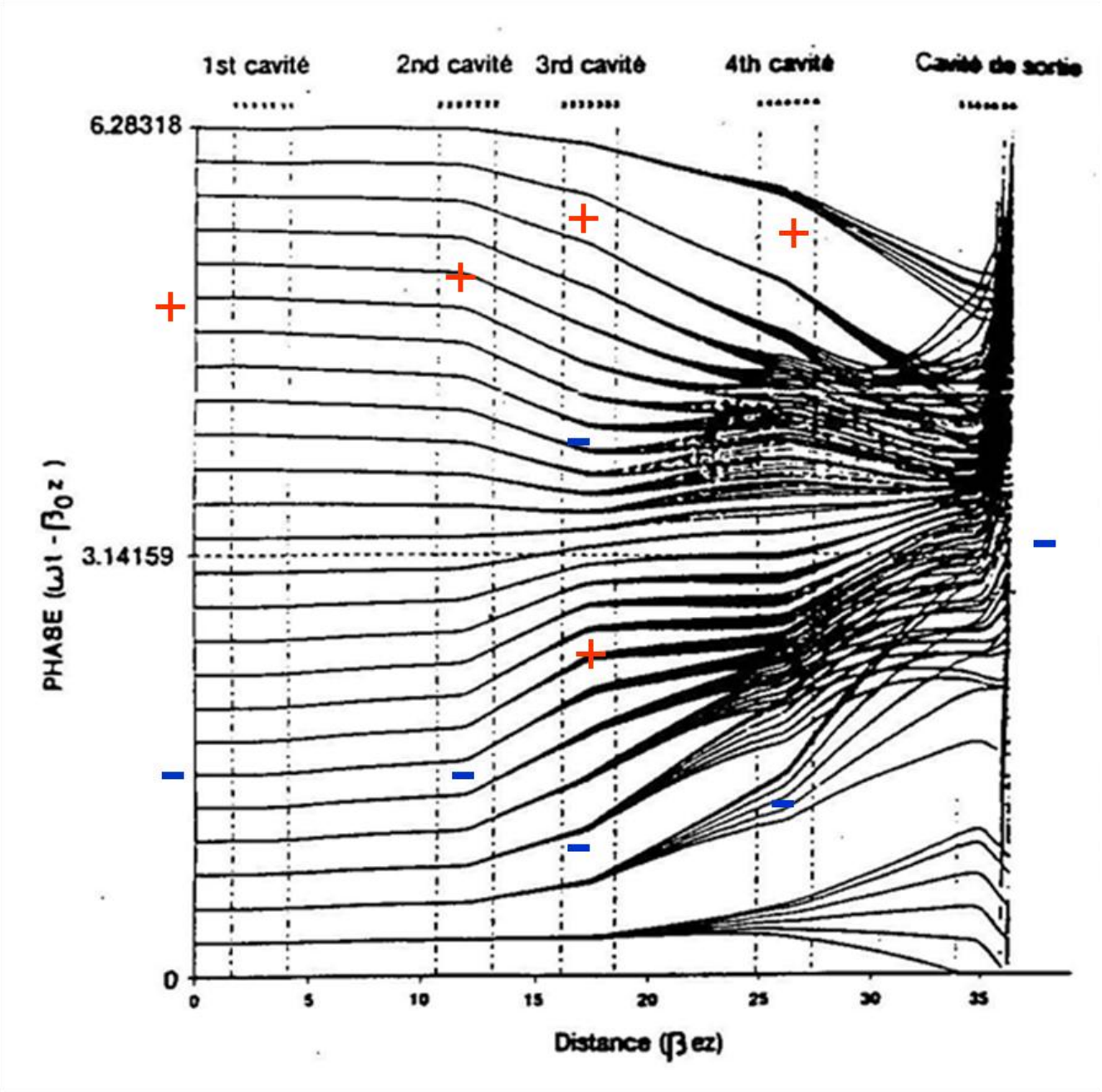

**Fig. 18:** Applegate diagram for a high efficiency klystron (Courtesy of Thales Electron Devices)

The tube illustrated has five cavities. The bunching produced by the first cavity is imperceptible on the scale of this diagram but it is sufficient to excite the r.f. fields in the second cavity. The second cavity is tuned to a frequency which is above the signal frequency so that it presents an inductive impedance to the beam current. As a result the bunch centre coincides with the neutral phase of the field in the cavity and further velocity modulation is added to the beam which produces much stronger bunching at the third cavity. The third cavity is tuned to the second harmonic of the signal frequency as can be seen from a careful examination of the diagram. The principal purpose of this cavity is to cause the electrons which lie farthest from the bunch centre to be gathered into the bunch. The use of a second harmonic cavity increases the efficiency of a klystron by at least ten percentage points. The splitting of the lines in the diagram which occurs at this plane is caused by a divergence in the behaviour of electrons in different radial layers within the electron beam. The fourth cavity is similar to the second cavity and produces still tighter bunching of the electrons. By the time they reach the final cavity nearly all the electrons are bunched into a phase range which is ± 90° with respect to the bunch centre. The output cavity is tuned to the signal frequency so that the electrons at the bunch centre experience the maximum retarding field and all electrons which lie within a phase range of ± 90° with respect to the bunch centre are also retarded. If the impedance of the output cavity is chosen correctly then a very large part of the kinetic energy of the bunched beam can be converted into r.f. energy. It should be noted that space-charge repulsion ensures that the majority of trajectories are nearly parallel to the axis at the plane of the output gap so that the kinetic energy of the bunch is close to that in the initial unmodulated beam.

### 5.2 Efficiency of klystrons

The output power of a klystron is given by

$$P_2 = \frac{1}{2} I_1 V_{eff} \tag{34}$$

where $I_1$ is the first harmonic r.f. beam current at the output gap and $V_{eff}$ is the effective output gap voltage. As in the case of the IOT the effective output gap voltage must be less than 90% of $V_0$ to ensure that the electrons have sufficient energy to leave the gap and enter the collector. In the IOT the peak current in the bunch cannot exceed the maximum instantaneous current available from the cathode and the maximum value of $I_1$ is approximately equal to half the peak current. In the klystron, however, the d.c. beam current is equal to the maximum current available from the cathode, and the bunches are formed by compressing the charge emitted in one r.f. cycle into a shorter period. In the theoretical limit the bunches become delta functions for which

$$I_1 = 2I_0 \,. \tag{35}$$

Thus the maximum possible value of $I_1$ in a klystron is four times that in an IOT with the same electron gun. The factor is actually greater than this because the current available from the triode gun in an IOT is less than that from the equivalent diode gun in a klystron. In practice the effects of space-charge mean that the limit given by Eq. (35) is not attainable but computer simulations have shown that the ratio $I_1/I_0$ can be as high as 1.6 to 1.7 at the output cavity. Then, by substitution in Eq. (34), we find that efficiencies of up to 75% should be possible.

It is to be expected that the maximum value of $I_1/I_0$ will decrease as the space-charge density in the beam increases. An empirical formula for the dependence of efficiency on beam perveance derived from studies of existing high-efficiency klystrons is given in Ref. [20]

$$\eta_e = 0.9 - 0.2 \times 10^{-6} K \,. \tag{36}$$

If it is assumed that the limit $K = 0$ corresponds to delta function bunches then it can be seen that Eq. (36) takes the maximum effective gap voltage to be $0.9\,V_0$.

The maximum efficiency of klystrons decreases with increasing frequency because of increasing r.f. losses and of the design compromises which are necessary. This is illustrated by Fig. 19 which shows the efficiencies of continuous-wave klystrons taken from manufacturers' data sheets. It should be emphasized that the performance of most of these tubes will have been optimized for factors other than efficiency.

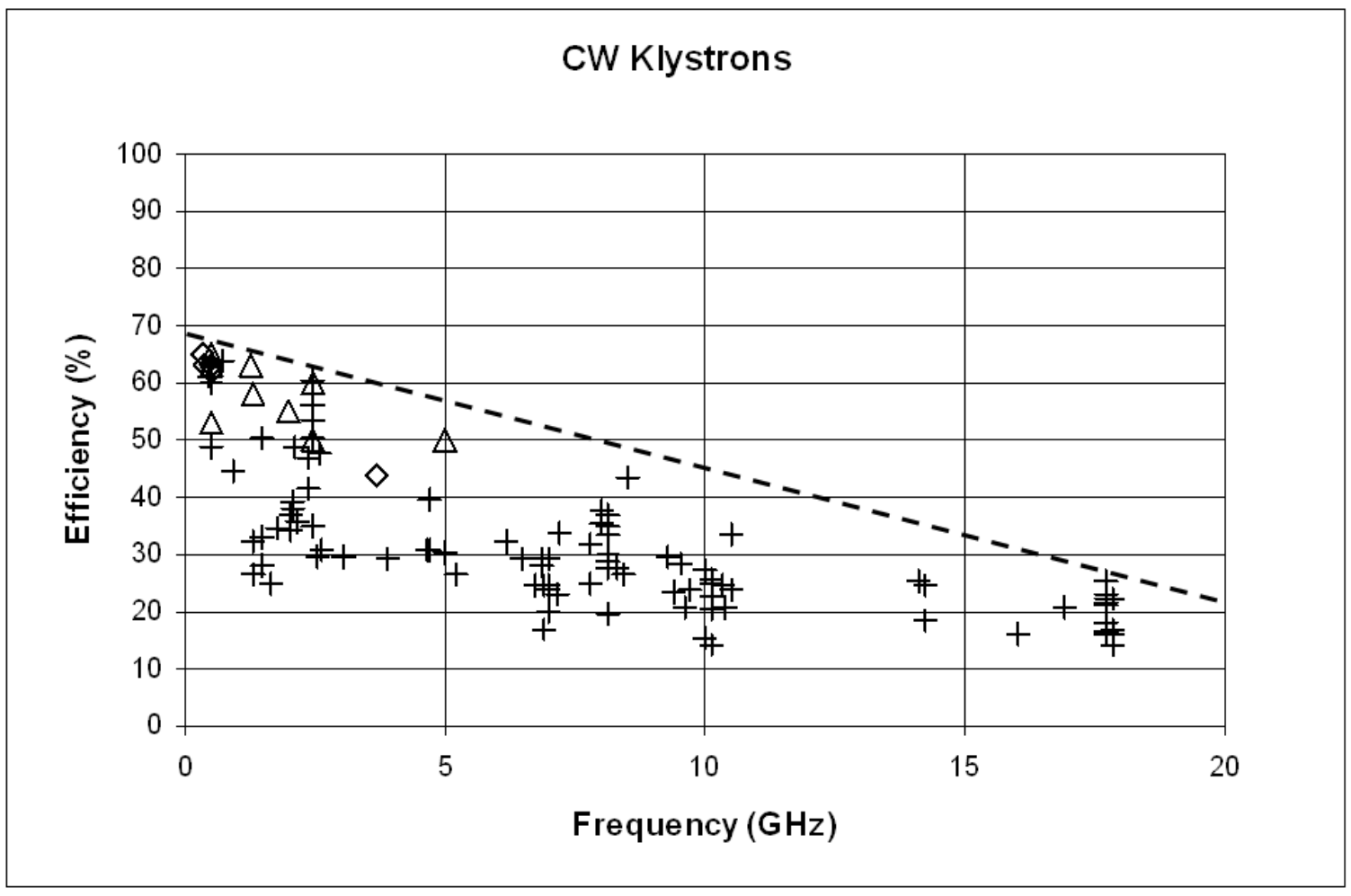

**Fig. 19:** Efficiencies of continuous-wave klystrons

### 5.3 Terminal characteristics of klystrons

The transfer characteristic of a klystron (Fig. 20) shows that the device is a linear amplifier at low signal levels but that the output saturates at high signal levels. A tube used in an accelerator would normally be run at, or close to, saturation to obtain the highest possible efficiency. The performance of a klystron is appreciably affected by variations in the beam voltage, signal frequency, and output match and we now examine these in turn.

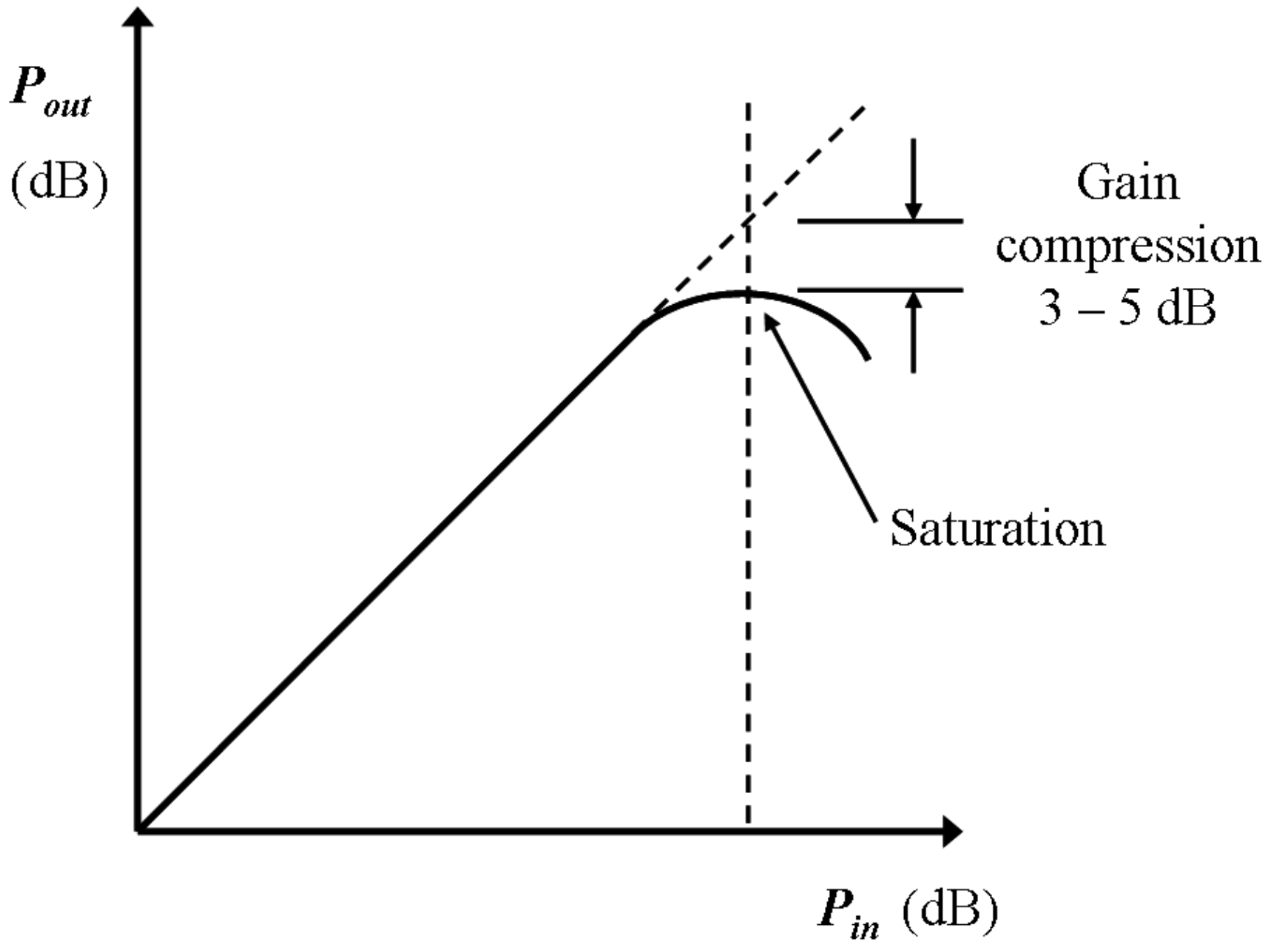

**Fig. 20:** Klystron transfer characteristics

Klystrons for use in accelerators are normally operated at or close to saturation. Figure 20 shows that the output power is then insensitive to variations of input power and, by extension, to variations of beam voltage. The effects on the phase of the output signal are more serious because of the distance from the input to the output. If the distance from the centre of the input gap to the centre of the output gap is $L$ then the phase difference between the input and the output is

$$\phi = \omega L / u_0 \tag{37}$$

where the beam velocity is given by

$$u_0 = c\left[1-\left(1+\left(eV/m_o c^2\right)\right)^{-2}\right]^{0.5} . \tag{38}$$

Thus if the normal beam voltage is 90 kV, the tube length is 1.17 m, and the frequency is 500 MHz the sensitivity of phase to changes in the beam voltage is -5.8 degrees per kV.

The transfer characteristic of a klystron with synchronously tuned cavities is essentially that of a resonant circuit as far as changes in frequency are concerned, namely

$$H(\omega) = \frac{R}{1 - jQ\left(\frac{\omega_0}{\omega} - \frac{\omega}{\omega_0}\right)} . \tag{39}$$

The effective $Q$ factor takes account of the combined effects of all the cavities and of any external loading. The klystron used in the example above has a bandwidth of 1 MHz giving an effective $Q$ factor of 500. Small changes in the centre frequency are produced by changes in the working temperature of the tube. If $\omega = \omega_0 + \delta\omega$ and if $\delta\omega$ is small, then

$$\text{Phase}\left(H\left(\omega\right)\right) = \arctan\left(-2Q\delta\omega / \omega_0\right) \tag{40}$$

giving a phase sensitivity of −63° per MHz. If the cavities are made from copper whose coefficient of thermal expansion is $16 \times 10^{-6}$ $K^{-1}$ then the sensitivity of phase to variations in temperature is 0.53° $K^{-1}$.

The output power and efficiency of a klystron are affected by the match of the load which is normally a circulator. This is usually represented by plotting contours of constant load power on a Smith chart of normalized load admittance. Figure 21 shows such a chart, known as a Rieke diagram, for a typical klystron. Care must be taken to avoid the possibility of voltage breakdown in the output gap. If the gap voltage becomes too high it is also possible for electrons to be reflected so reducing the efficiency of the tube and providing a feedback path to the other cavities which may cause the tube to become unstable. The forbidden operating region is shown by shading on the diagram. A further complication is provided by the effect of harmonic signals in the output cavity. Since the klystron is operated in the non-linear regime to obtain maximum efficiency, it follows that the signal in the output waveguide will have harmonic components. These are incompletely understood but it is known that the reflection of harmonic signals from external components such as a circulator can cause the klystron output to behave in unexpected ways.

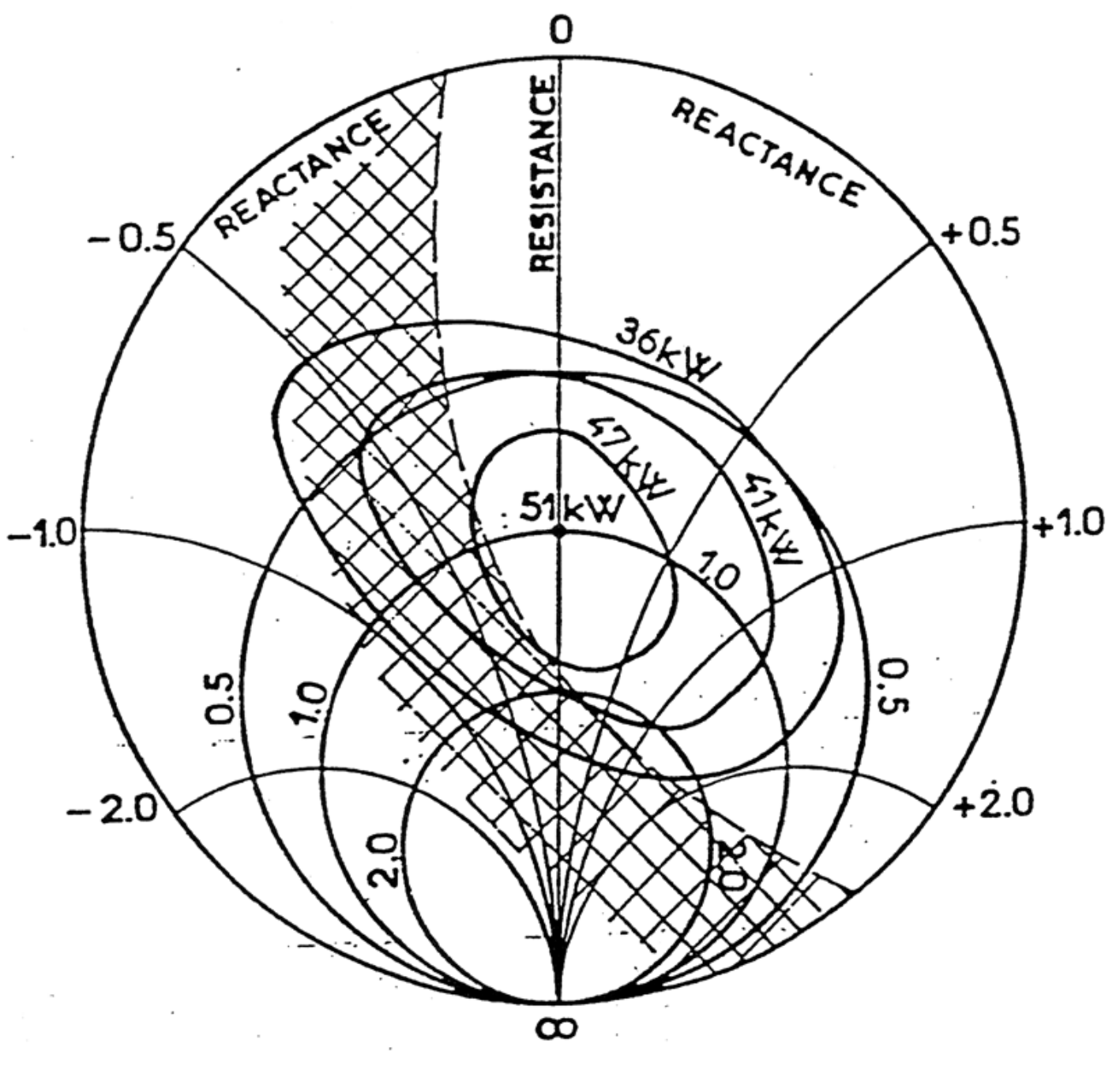

**Fig. 21:** Rieke diagram for a klystron (Courtesy of Thales Electron Devices)

## 5.4 Typical klystrons

### *5.4.1 UHF television klystrons*

At moderate powers in the UHF band it is possible to use klystrons designed for use in UHF television transmitters as power sources for accelerators. Tubes are available with output powers in the range 10 kW to 70 kW and gains of 30 dB to 40 dB. The beam current can usually be controlled independently of the beam voltage by the voltage applied to a separate modulating anode. The conversion efficiency is around 50% but in some modern tubes this is increased by the use of a multi-element depressed collector. This type of collector has several electrodes at different potentials between ground and cathode potentials. The principle of operation is illustrated by the two-stage collector shown in Fig. 22 in which the beam current is collected either by the tube body or by a depressed electrode. The d.c. power supplied to the tube is given by

$$P_{DC} = I_C\left(V_0 - V_C\right) + I_b V_0 = I_0 V_0 - I_C V_C \tag{41}$$

where the symbols are defined in Fig. 22. From Eq. (2) the efficiency of the tube is

$$\eta_e = \frac{P_{RF}}{I_0 V_0 - I_C V_C} . \tag{42}$$

Since the denominator in Eq. (42) is less than would be the case if the collector were not depressed, the overall efficiency of the tube is increased by collector depression. Because the electrons strike the collector electrodes with reduced energy, the power dissipated in the collector is reduced [21]. The number of electrodes can be as great as ten but is usually limited to three or four by practical considerations. This technique is not used with very high power tubes because of the difficulty of cooling electrodes which are at a high voltage with respect to ground.

If greater power is required than can be supplied by one tube it is possible to operate several tubes in parallel. A 450 kW, 800 MHz amplifier for the CERN SPS comprises eight modified UHF television klystrons operated in parallel.

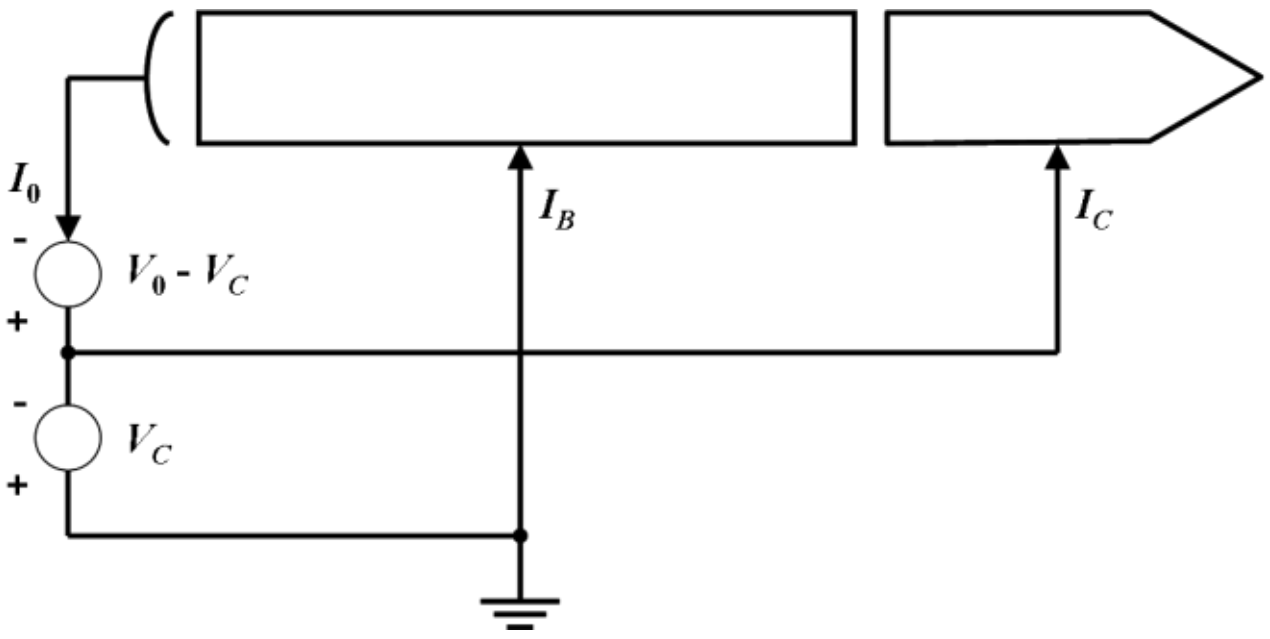

**Fig. 22:** Circuit diagram of a tube with a depressed collector

### *5.4.2 Super-power klystrons*

Klystrons which have been developed specifically for use in accelerators are commonly known as super-power klystrons. Tables 7 and 8 summarize the state of the art for these tubes. The beam voltage is limited by the need to avoid voltage breakdown in the electron gun. It can be seen from the tables that the typical beam voltages are higher for pulsed tubes than for continuous-wave tubes because the breakdown voltage is higher for short pulses than for steady voltages. The beam current is limited by the current density which is available at the cathode and by the area of the cathode which decreases with frequency. The saturation current density of thermionic cathodes is greater for short (microsecond) pulses than for d.c. operation.

**Table 7:** Characteristics of typical continuous-wave super-power klystrons

| | **TH 2089** | **VKP-7952** | **TH 2103C**[a] |
|---|---|---|---|
| Manufacturer | Thales | CPI | Thales |
| Frequency (MHz) | 352 | 700 | 3700 |
| Beam voltage (kV) | 100 | 95 | 73 |
| Beam current (A) | 20 | 21 | 22 |
| RF output power (MW) | 1.1 | 1.0 | 0.7 |
| Gain (dB) | 40 | 40 | 50 |
| Efficiency (%) | 65 | 65 | 44 |

a. This tube was developed for heating plasmas for nuclear fusion experiments.

**Table 8:** Characteristics of typical pulsed super-power klystrons

| | Ref. [22] | Ref. [23] | Ref. [24] |
|---|---|---|---|
| Frequency (GHz) | 2.87 | 3.0 | 11.4 |
| Pulse length (μs) | 1.0 | 1.0 | 1.6 |
| Beam voltage (kV) | 475 | 610 | 506 |
| Beam current (A) | 620 | 780 | 296 |
| RF output power (MW) | 150 | 213 | 75 |
| Gain (dB) | 59 | 58 | 60 |
| Efficiency (%) | 51 | 44 | 50 |

## 5.5 Multiple-beam klystrons

We have seen that the efficiency of a klystron is determined by the perveance of the electron beam so that, to get high efficiency, it is necessary to use a high-voltage, low-current beam. The use of high voltages produces problems with voltage breakdown and it is therefore difficult to obtain very high power with high efficiency. One solution to this problem is to use several electron beams within the same vacuum envelope as shown in Fig. 23. A klystron designed in this way is known as a Multiple-Beam Klystron (MBK). The individual beams have low perveance to give high efficiency whilst the output power is determined by the total power in all the beams. The principle of the MBK has been known for many years Ref. [25] but, until recently, the only such tubes constructed were in the former Soviet Union for military applications. The first MBK designed specifically for use in particle accelerators was the Thales type TH1801 whose performance is shown in Table 9, see Ref. [26].

**Table 9:** Characteristics of a multiple-beam klystron

| | *TH 1801* |
|---|---|
| Frequency | 1300 MHz |
| Beam voltage | 115 kV |
| Beam current | 133 A |
| Number of beams | 7 |
| Power | 9.8 MW |
| Pulse length | 1.5 ms |
| Efficiency | 64% |
| Gain | 47 dB |

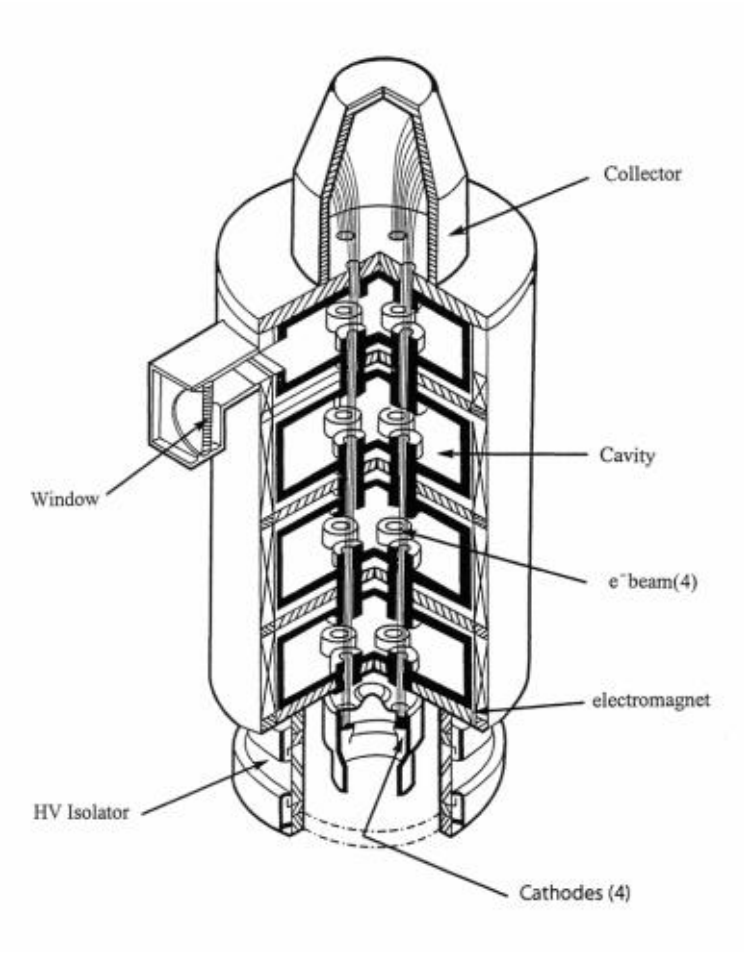

**Fig. 23:** Arrangement of a multiple-beam klystron (Courtesy of Thales Electron Devices)

## 6 Magnetrons

The principle of operation of the magnetron is illustrated in Fig. 24. The tube has a concentric cylindrical geometry. Electrons emitted from the cathode are drawn towards the surrounding anode by the potential difference between the two electrodes. The tube is immersed in a longitudinal magnetic field which causes the electron trajectories to become cycloidal so that, in the absence of any r.f. fields, the diode is cut off, no current flows, and the electrons form a cylindrical space-charge layer around the cathode. The anode is not a smooth cylinder but carries a number of equally spaced vanes such that the spaces between them form resonant cavities. The anode supports a number of resonant modes with azimuthal r.f. electric field. The one used for the interaction is the $\pi-$mode in which the fields in adjacent cavities are in anti-phase with one another. The r.f. fields in the anode are initially excited by electronic noise and there is a collective interaction between the fields and the electron cloud which causes some electrons to be retarded. These electrons move outwards forming 'spokes' of charge whose number is half the number of the cavities in the anode. The spokes rotate in synchronism with the r.f. field of the anode and grow until electrons reach the anode and current flows through the device. The electron velocities are almost constant during the interaction and the energy transferred to the r.f. field comes from their change in potential energy. The magnetron is an oscillator whose power output grows until it is limited by non-linearity in the interaction. RF power is extracted from the anode via a coupler and vacuum window.

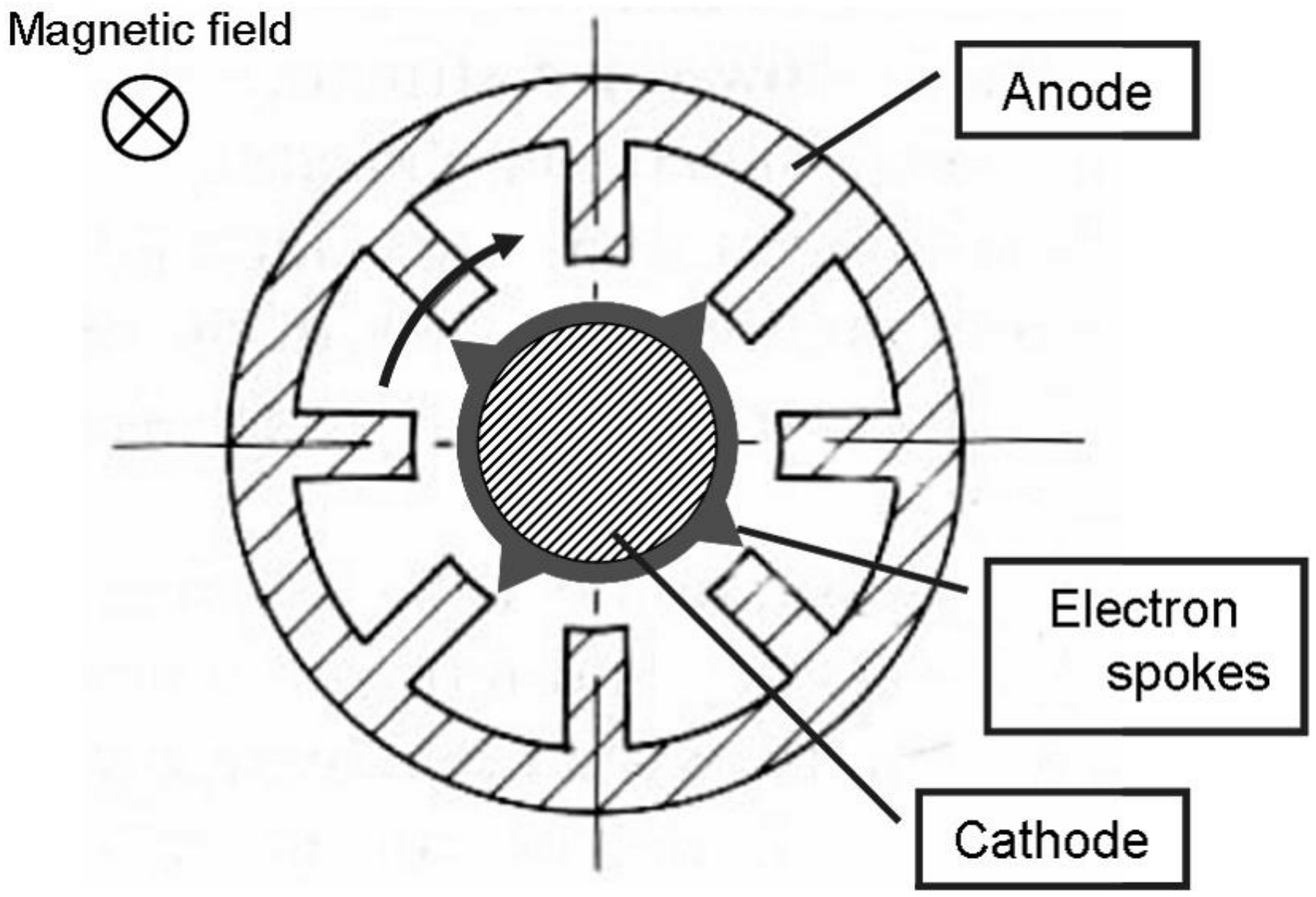

**Fig. 24:** Arrangement of a magnetron oscillator

The magnetron is a compact device which is capable of achieving efficiencies of up to 90% and it has been recognized for many years that it would be an attractive alternative to other tubes for powering particle accelerators. However, because it is a free-running oscillator, the frequency is not stable enough for use in most accelerators. The frequency of a magnetron varies with the current flowing through the tube (known as frequency pushing) and it is possible to use this to provide a degree of control. On its own this is not sufficient. It is also possible to lock the phase of a free-running oscillator by injecting r.f. power at the desired frequency. The power required increases with the difference between the natural frequency and the locked frequency and it is found that the power required to lock the phase of a magnetron is typically about 10% of the output power of the tube. This power is unacceptably high. Recent work has shown that, when the frequency of a magnetron is first stabilised by a control loop using frequency pushing, it is then possible to lock the phase with an injected r.f. signal which is less than 0.1% of the output power of the tube [27]. Thus locked magnetrons may be used in the future for powering accelerators [28].

### 6.1 Medical linac magnetrons

Notwithstanding the stability problems mentioned above, magnetrons have been used for many years to provide the r.f. power for linear accelerators used for radiotherapy and industrial radiography. The performance of a tube of this kind is shown in Table 10. Note that the anode voltage is considerably less than would be required for a klystron with similar performance.

**Table 10:** Performance of a magnetron for medical linacs

| | **MG 5350** |
|---|---|
| Manufacturer | e2v technologies |
| Frequency | 2855 MHz |
| Beam voltage | 51 kV |
| Beam current | 240 A |
| Power | 5.5 MW |
| Pulse length | 2.3 μsec |
| Duty | 0.00055 |
| Efficiency | 45% |

## 7 Gyrotrons

An alternative type of tube for producing very high pulsed r.f. power at high frequencies is the gyrotron. This type of tube has been the subject of intensive developmental work mainly with a view to providing r.f. power for plasma heating experiments. A good summary of this work and of the development of other, novel, high-power r.f. sources is given in Ref. [29]. The gyrotron employs the interaction between an annular electron beam and the azimuthal electric field of a circular waveguide mode as shown in Fig. 25. There is a strong axial magnetic field so that the electrons move in small orbits at the cyclotron frequency within the beam as shown. The cyclotron frequency is made equal to the signal frequency. At frequencies above 60 GHz this means that a superconducting solenoid is needed to produce the magnetic field. It is essential to the working of the gyrotron that the electrons have relativistic velocities. The cyclotron frequency is then a function of the electron velocity given by

$$\omega_c = \left(eB/m_0\right)\left(1 - v^2/c^2\right)^{0.5} \tag{52}$$

where $B$ is the magnetic flux density, $e/m_0$ is the charge to mass ratio of the electron, $v$ is the electron velocity, and $c$ is the velocity of light. As the electron velocity increases, the cyclotron frequency decreases so that the faster electrons lag behind the r.f. field. Slow electrons, similarly, lead the field and phase bunching occurs with a net transfer of energy to the r.f. field.

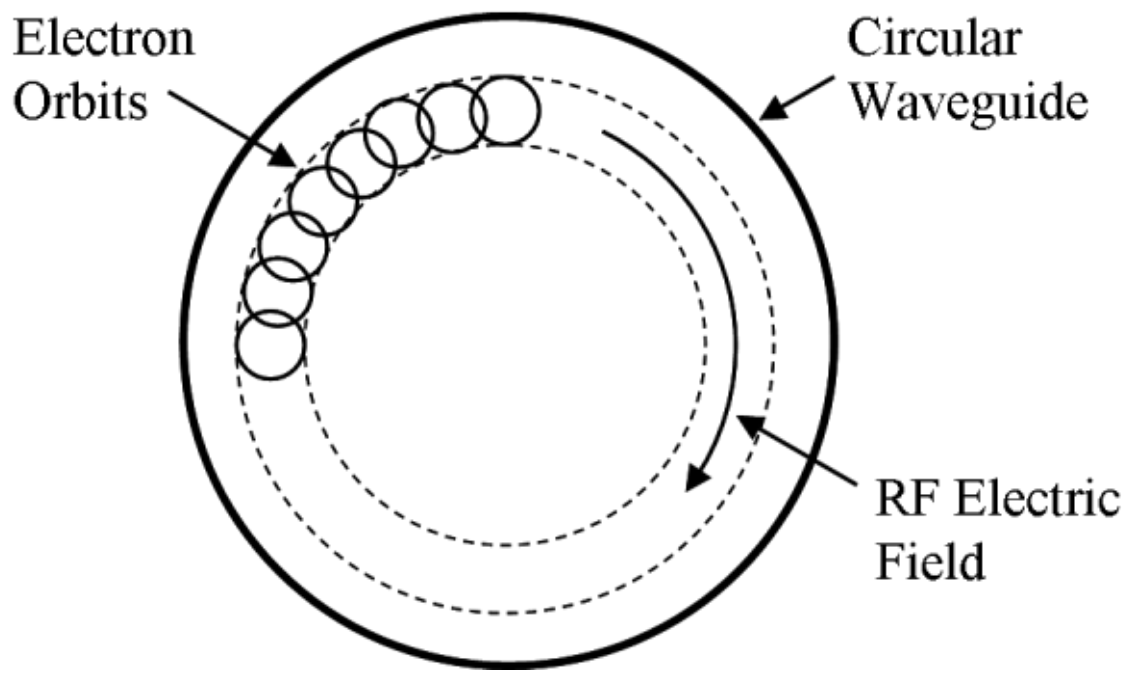

**Fig. 25:** Principle of the gyrotron interaction

Commercially available gyrotrons are oscillators with the general arrangement shown in Fig. 26. The hollow electron beam is produced by a magnetron electron gun and confined by an axial magnetic field. The interaction takes place with a TE mode in a section of cylindrical waveguide which is made resonant by the mismatches at its ends. The modes in the waveguide have phase velocities greater than the velocity of light and the electric field is not confined to the region close to the metallic surface as is the case in a klystron. It is therefore possible to make the transverse dimensions of the beam and the waveguide bigger than in a klystron operating at the same frequency. The transverse dimensions can also be increased by operating in a higher order mode of the waveguide. The spent electron beam is allowed to spread sideways so that it is collected on the wall of the larger, cylindrical output waveguide. The r.f. output power passes down this guide and through the output window. At millimetre wavelengths it is usual to use an over-moded output waveguide to avoid problems with breakdown in it. Table 11 shows the characteristics of some typical gyrotron oscillators. Because these tubes are oscillators they suffer from the same drawbacks as magnetrons for accelerator applications. Their efficiencies are generally lower than those of comparable klystrons but they have the potential for use at frequencies greater than 12 GHz where the development of high-power klystrons is difficult.

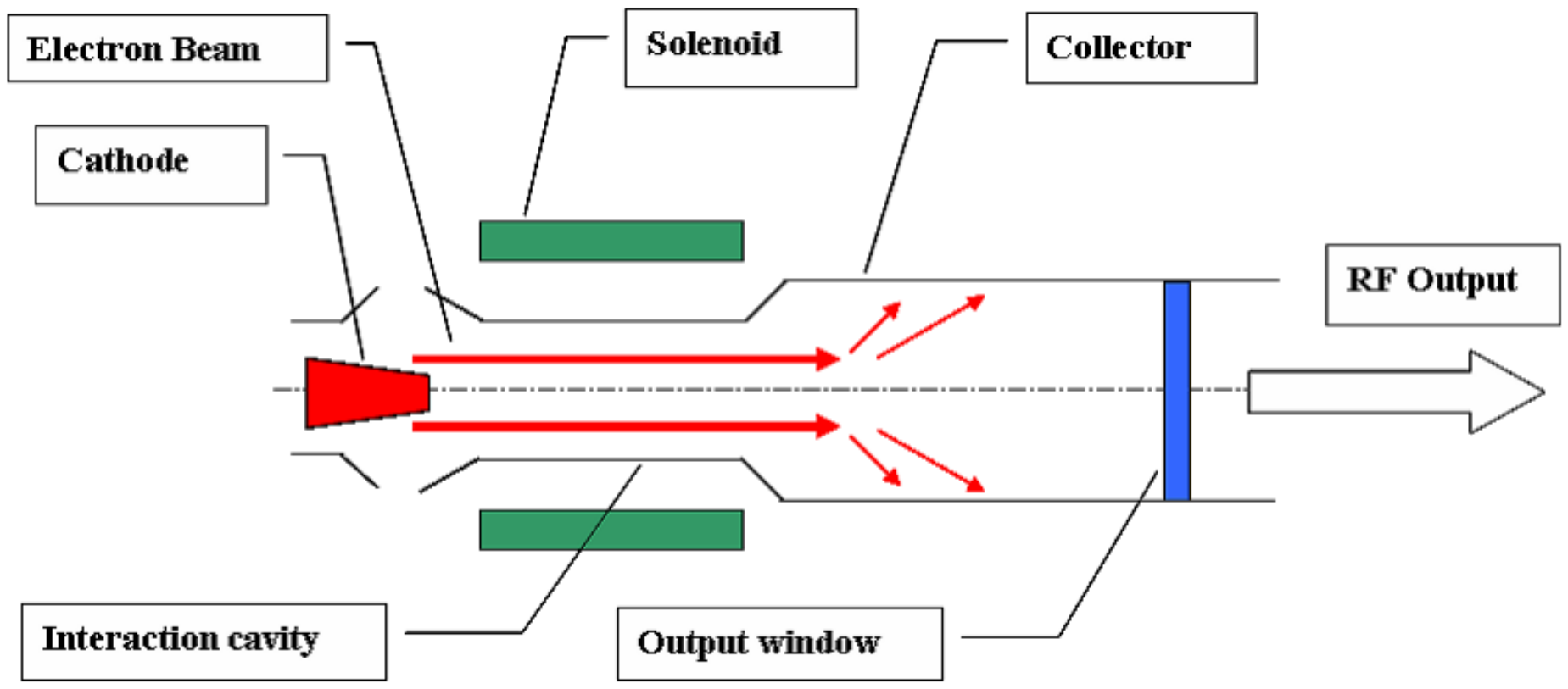

**Fig. 26:** Arrangement of a gyrotron oscillator

**Table 11:** Characteristics of typical pulsed gyrotron oscillators

| | **TH1504** | **TH1506A** |
|---|---|---|
| Manufacturer | Thales | Thales |
| Frequency (GHz) | 8 | 110 |
| Pulse length (s) | 1 | 5 |
| Duty | 1/600 | 1/3 |
| Beam voltage (kV) | 85 | 85 |
| Beam current (A) | 27 | 22 |
| Power (MW) | 1 | 0.5 |
| Efficiency (%) | 41 | 30 |

## 7.1 Gyro-TWT amplifiers

Experimental gyrotron amplifiers which are analogues of klystrons and travelling-wave tubes have been built for some years but these have not yet found application in accelerators [30]. A recent development is the gyro-TWA illustrated in Fig. 27 which employs a helical waveguide [31]. The characteristics of this tube are shown in Table 12.

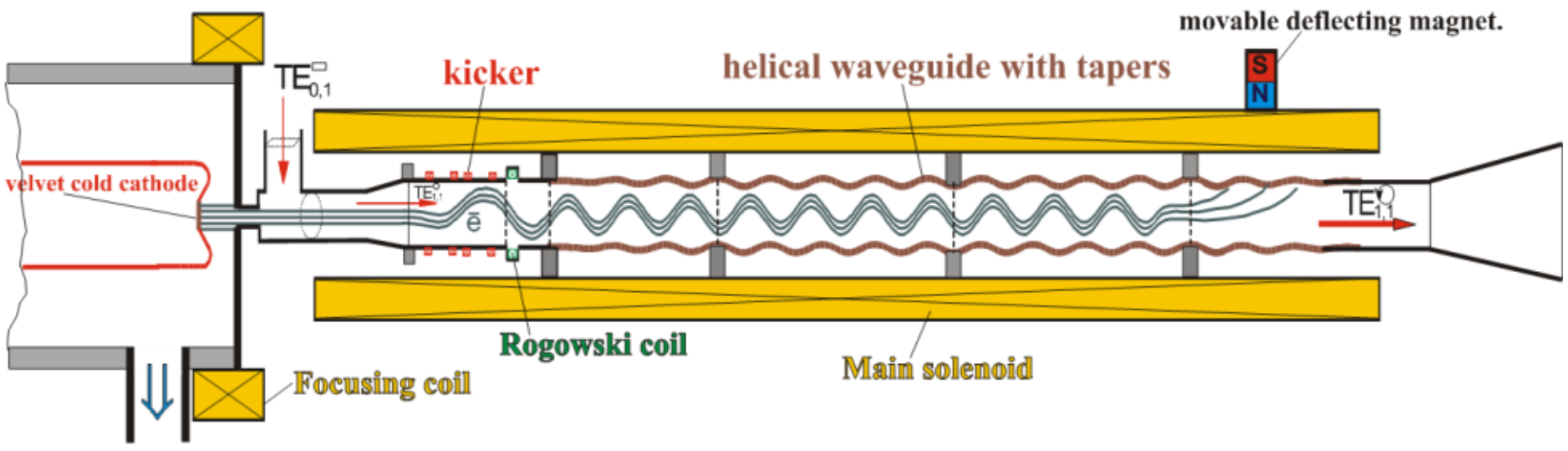

**Fig. 27:** Arrangement of a gyro-TWA amplifier (Courtesy of University of Strathclyde)

**Table 12:** Characteristics of a gyro-TWA amplifier

| | |
|---|---|
| Frequency | 8.4–10.4 GHz |
| Pulse length | 1.0 µs |
| Beam voltage | 185 kV |
| Beam current | 6 A |
| Power | 220 kW |
| Gain | 24 dB |
| Efficiency | 20% |

## 8 Limitations of vacuum tubes

The performance of high-power vacuum tubes is limited by a number of factors which operate in much the same way for all devices. The chief of these are heat dissipation, voltage breakdown, output window failure, and multipactor discharges.

The dimensions of the r.f. structures and the windows of microwave tubes generally scale inversely with frequency. The maximum continuous, or average, power which can be handled by a particular type of tube depends upon the maximum temperature which the internal surfaces can be allowed to reach. Now this temperature is independent of the frequency so the power which can be dissipated per unit area is constant. Gyrotrons can handle a higher power than klystrons of the same frequency because they have simpler structures and, if operated in a higher order mode, ones which are larger for a given frequency.

The power is also limited by the power in the electron beam. The beam diameter scales inversely with frequency and the beam current density is determined by the maximum attainable magnetic focusing field. Since that field is independent of frequency, the beam current scales inversely with the square of the frequency. The beam voltage is related to the current by the gun perveance which usually lies in the range 0.5 to 2.0 for power tubes. The maximum gun voltage is limited by the breakdown field in the gun and so varies inversely with frequency for constant perveance. These considerations suggest that the maximum power obtainable from a tube of a particular type varies as frequency to the power −2.5 to −3.0 depending upon the assumptions made. For pulsed tubes the peak power is limited by the considerations in this paragraph and the mean power by those in the preceding one.

The efficiencies of tubes tend to fall with increasing frequency. This is partly because the r.f. losses increase with frequency and partly because of the design compromises which must be made at higher frequencies.

The maximum power obtainable from a pulsed tube is often determined by the power-handling capability of the output window. Very high power klystrons commonly have two windows in parallel to handle the full output power. Windows can be destroyed by excessive reflected power, by arcs in the output waveguide, by X-ray bombardment, and by the multipactor discharges described in the next section. The basic cause of failure is overheating and it is usual to monitor the window temperature and to provide reverse power and waveguide and cavity arc detectors.

### 8.1 Multipactor discharge

Multipactor is resonant radiofrequency vacuum discharge which is sustained by secondary electron emission [32]. Consider a pair of parallel metal plates in vacuum with a sinusoidally varying voltage between them. If an electron is liberated from one of the plates at a suitable phase of the r.f. field it will be accelerated towards the other plate and may strike it and cause secondary electron emission. If the phase of the field at the moment of impact is just 180 degrees from that at the time when the electron left the first plate then the secondary electrons will be accelerated back towards the first plate. These conditions make it possible for a stable discharge to be set up if the secondary electron emission coefficients of the surfaces are greater than unity. It is found that phase focusing occurs so that electrons which are emitted over a range of phases tend to be bunched together.

The secondary emission coefficients of many materials vary with the energy of electrons at normal incidence according to the universal curve shown in Fig. 28 where δ is the secondary electron emission coefficient and $E_p$ is the energy of the incident primary electron. The constants of this curve for a number of materials used in vacuum tubes are given in Table 13.

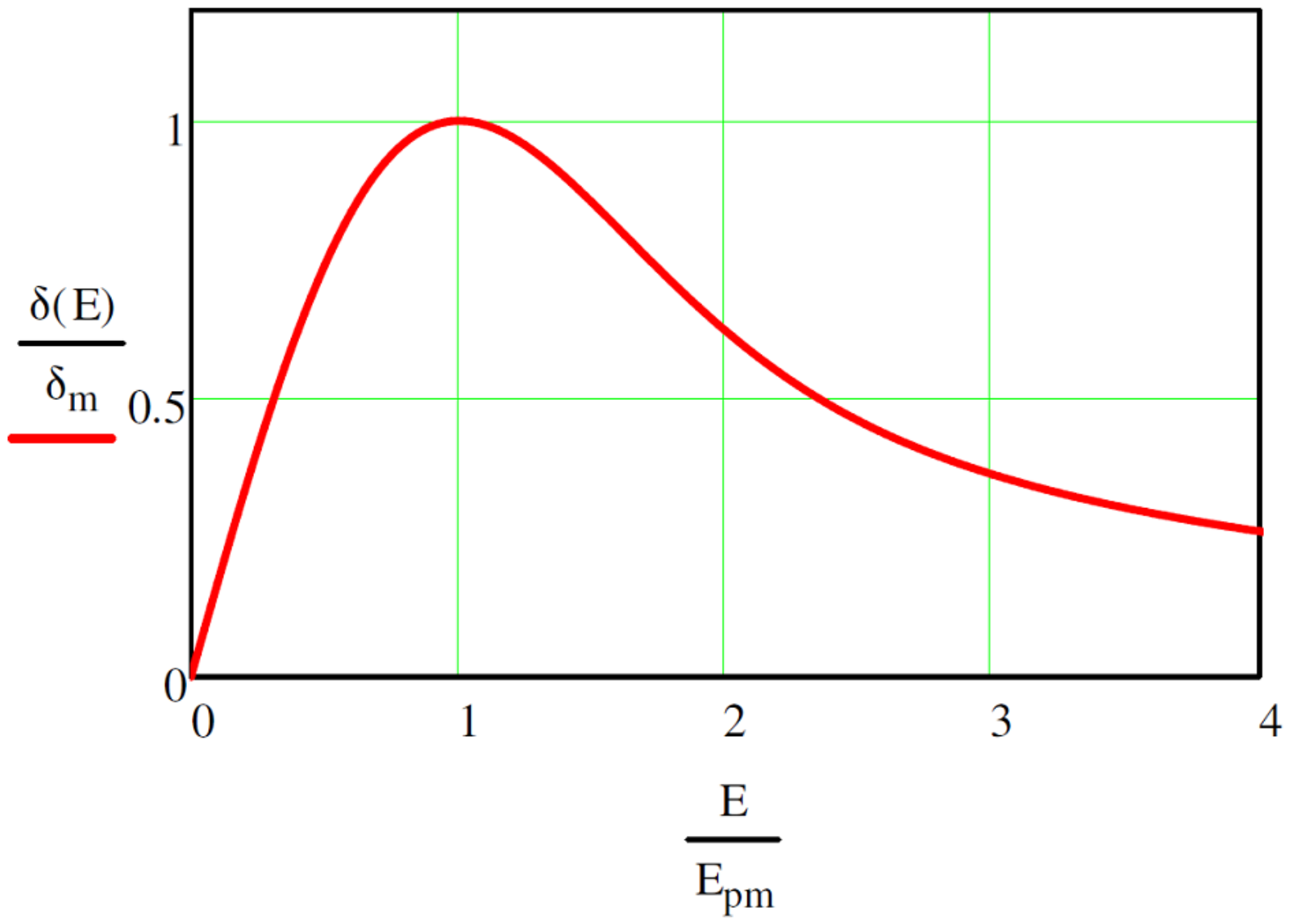

**Fig. 28:** Universal secondary electron emission curve

**Table 13:** Secondary electron emission coefficients of materials used in vacuum tubes

| **Material** | $\delta_m$ | $E_{pm}$ **(V)** |
|---|---|---|
| Copper | 1.3 | 600 |
| Platinum | 1.8 | 800 |
| Carbon | 0.45 | 500 |
| Alumina | 2.35 | 500 |

In order for a two-surface multipactor discharge to be sustained it is necessary both that the secondary electron emission coefficient be greater than unity and that the r.f. voltage between the electrodes produces impact energies in the range for which this occurs. The discharge can, therefore, only occur within a fairly limited range of voltages and products of frequency and electrode separation as illustrated by Fig. 29 where $f$ is the frequency, $d$ the separation of the plates, and $V_1$ the amplitude of the r.f. voltage between them. The limits in the vertical direction are set by the need for the secondary electron emission coefficient to be greater than unity. Table 13 shows that this happens for a small range of energies which are of the order of a few hundred electronvolts. The limits in the horizontal direction are set by the need for the correct phase relationships to exist. Figure 29 also shows the ranges in which higher-order multipactor discharges can occur where the electrons cross the gap in an odd number of half r.f. cycles. The two-surface multipactor discharge typically involves currents of less than 1 A and voltages of a few hundred volts so the power is moderate and the discharge is not normally destructive. It is probable that discharges of this kind occur in most microwave power tubes and their main effect is to cause some additional loss, noise, and loading of the r.f. circuit.

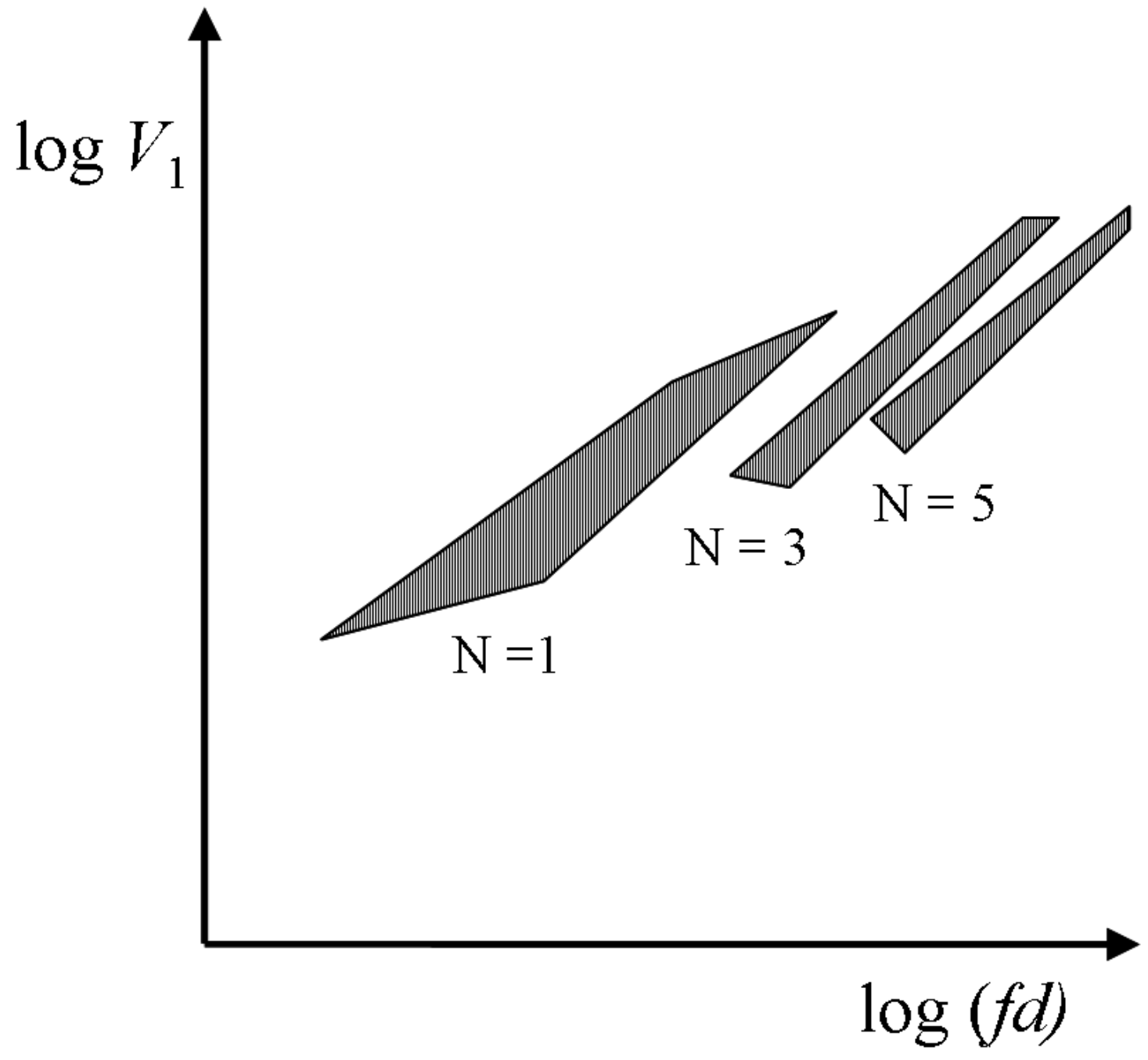

**Fig. 29:** Typical regions where multipactor can occur (After Vaughan [32], © IEEE 1988, reproduced with permission)

A more serious kind of multipactor discharge can occur in the presence of a magnetic field. The electron trajectories are bent by the field so that the impacts made on the surfaces are oblique. This type of discharge can involve impacts on either one or two surfaces. When the primary electrons strike a surface obliquely the peak value of δ is greater than for normal incidence, and the range of energies over which δ is greater than unity is greatly increased. Thus crossed-field multipactor discharges can occur at much higher energies than the simple multipactor and they are, in consequence, potentially much more damaging. Because strong magnetic fields are used to focus linear-beam tubes it is quite possible for the conditions for crossed-field multipactor to exist somewhere within the tube. The manufacturer will normally have taken steps to ensure that this is not the case but, if the magnetic field around the tube is disturbed in any way, by the field of a circulator for example, then it is possible for a destructive discharge to occur.

It is also possible for multipactor discharges to occur on ceramic surfaces with surface charge providing a static field. It should be noted that this type of discharge is not resonant and does not require the presence of an r.f. electric field. The local heating of a window ceramic in this way can be sufficient to cause window failure.

Signs of multipactor are heating, changed r.f. performance, window failure, and light and X-ray emissions. Multipactor can sometimes be suppressed by changing shape of surfaces, by surface coatings, and by the imposition of static electric and magnetic fields.

## 9 Cooling and protection

### 9.1 Cooling power tubes

The power tubes used in accelerators typically have efficiencies between 40% and 70%. It follows that a proportion of the d.c. input power is dissipated as heat within the tube. The heat to be dissipated is between 40% and 150% of the r.f. output power provided that the tube is never operated without r.f. drive. If a linear beam tube is operated without r.f. drive then the electron collector must be capable of dissipating the full d.c. beam power. The greater part of the heat is dissipated in the anode of a tetrode or in the collector of a linear-beam tube. These electrodes are normally cooled in one of three ways: by blown air (at low power levels), by pumped liquid (usually de-ionized water), or by vapour phase cooling. The last of these may be less familiar than the others and needs a little explanation.

The electrode to be cooled by vapour phase cooling is immersed in a bath of the liquid (normally de-ionized water) which is permitted to boil. The vapour produced is condensed in a heat exchanger which is either within the cooling tank (see Fig. 30) or part of an external circuit. The cooling system therefore forms a closed loop so that water purity is maintained. In all water-cooling systems it is important to maintain the water purity to ensure that the electrodes cooled are neither contaminated nor corroded. Either of these effects can degrade the effectiveness of the cooling system and cause premature failure of the tube. In blown-air systems careful filtering of the air is necessary for the same reasons.

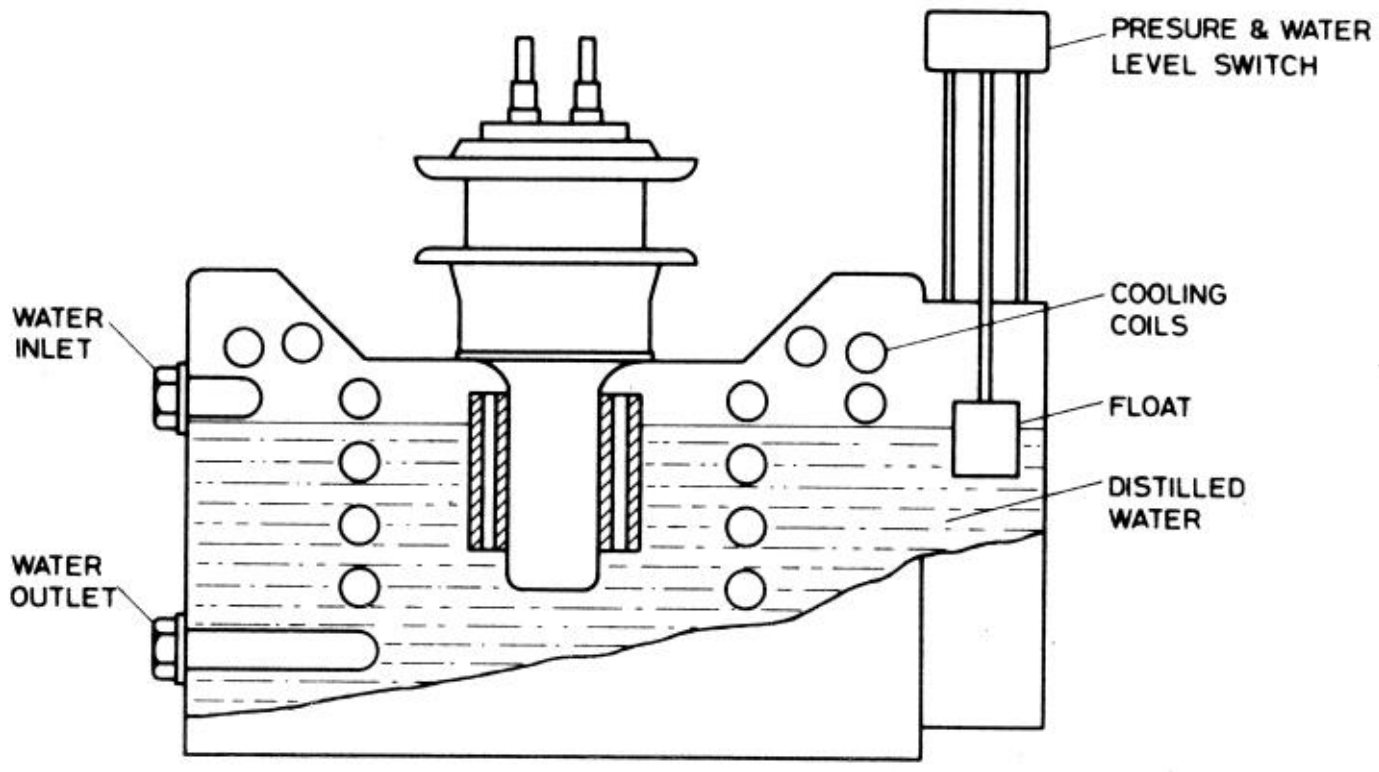

**Fig. 30:** Vapour phase cooling of a tetrode (Courtesy of e2v technologies)

It is important to remember that, in a high-power tube, appreciable quantities of heat may be dissipated on parts of the tube other than the anode or collector especially if a fault occurs during operation. It is common to provide air or water cooling for these regions also. Inadequate cooling may lead to the internal distortion or melting of the tube and its consequent destruction. Further information on the cooling of tubes is given in Refs. [11, 33].

### 9.2 Tube protection

Power tubes are very expensive devices and it is vital that they be properly protected when in use. The energy densities in the tubes and their power supplies are so high that it is easy for a tube to be destroyed if it is not properly protected. However, with adequate protection, tubes are in fact very good at withstanding accidental overloads and may be expected to give long, reliable service.

Two kinds of protection are required. First a series of interlocks must be provided which ensures that the tube is switched on in the correct sequence. Thus it must be impossible to apply the anode voltage until the cathode is at the correct working temperature and the cooling systems are functioning correctly. The exact switch-on sequence depends upon the tube type and reference must be made to the manufacturer’s operating instructions. The sequence must also be maintained if the tube has to be restarted after tripping off for any reason.

The second provision is of a series of trips to ensure that power is removed from the tube in the event of a fault such as voltage breakdown or excessive reflected power. Again the range of parameters to be monitored and the speed with which action must be taken varies from tube to tube. Examples are: coolant flow rate, coolant temperature, tube vacuum, output waveguide reverse power, and electrode over-currents. If a tube has not been used for some time it is sometimes necessary to bring it up to full power gradually to avoid repeated trips. The manufacturer’s operating instructions should be consulted about this. If a tube trips out repeatedly it is wise to consult the manufacturer to avoid the risk of losing it completely by unwise action taken in ignorance of possible causes of the trouble. General information about tube protection and safe operation is given in Refs. [11, 33].

## 10 The SLAC energy doubler

Sometimes higher pulsed powers are required for accelerators than can be obtained from a single tube. One way of achieving this is the energy doubler devised at SLAC (SLED) [34]. The principle of operation of this device is illustrated by Fig. 31. The input to the klystron is via a 180° switched phase shifter. The output passes to a 3 dB coupler connected to a pair of cavities. If the cavity fields are initially zero then part of the output power of the klystron goes into building them up whilst the remainder is reflected and passes on towards the accelerator. As the cavity fields increase they re-radiate power in anti-phase with the incident signal so that the mean power transmitted to the accelerator is small. The cavities are over-coupled so that the peak re-radiated power rises to a level greater than the transmitted power. If the phase of the klystron drive is then abruptly reversed the re-radiated and transmitted signals are in phase with each other and the power transmitted to the accelerator rises steeply as shown. The power level then decays as the energy stored in the cavities is discharged. The process can be repeated to provide a train of very high energy pulses. The principle is similar to that of a pulse modulator with r.f. storage in the cavities replacing d.c. storage in capacitors. The original SLED system used a pulsed 30 MW klystron to produce SLED output pulses at 125 MW. A similar system has been suggested which employs a continous-wave klystron [34].

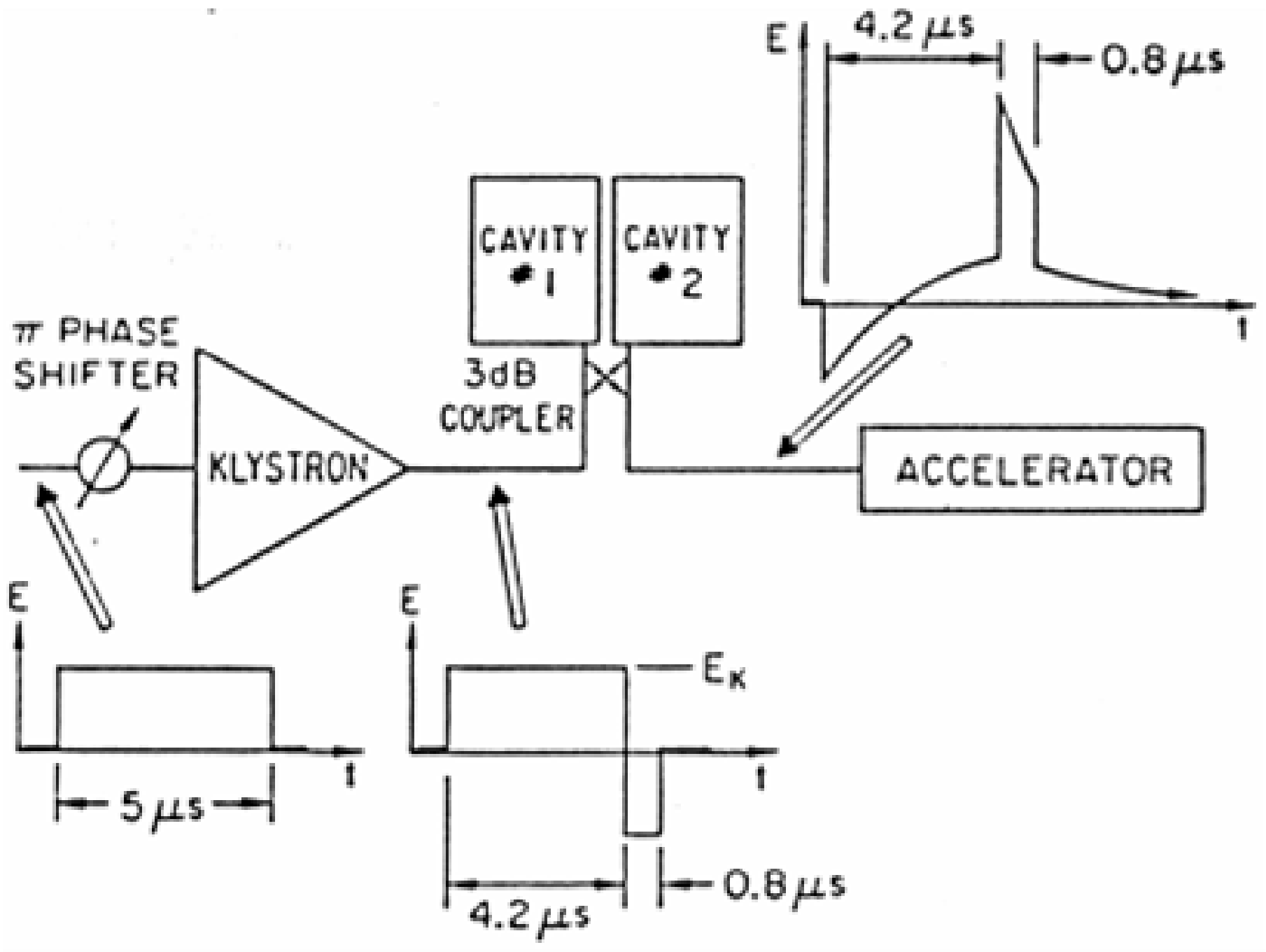

**Fig. 31:** The SLAC energy doubler (© IEEE 1975, reproduced with permission)

## 11 Conclusion

This paper has provided an introduction to the main types of r.f. power source which are, or may be, used in particle accelerators. Figure 32 shows the state of the art as a function of frequency for the r.f. power sources currently used in accelerators. Solid-state sources can compete with tubes at the lower frequencies and power levels and are likely to become more commonly used. The fall-off in power output at high frequencies for each type of tube is related to the fundamental principles of its operation as discussed in Section 8. The power achieved by klystrons at low frequencies does not generally represent a fundamental limitation but merely the maximum which has been demanded to date. For tetrodes and solid-state devices the maximum power is probably closer to the theoretical limits for those devices. In any case higher powers can be produced by parallel operation.

For further information on the theory of microwave tubes and for suggestions for background reading see Refs. [30, 33, 36].

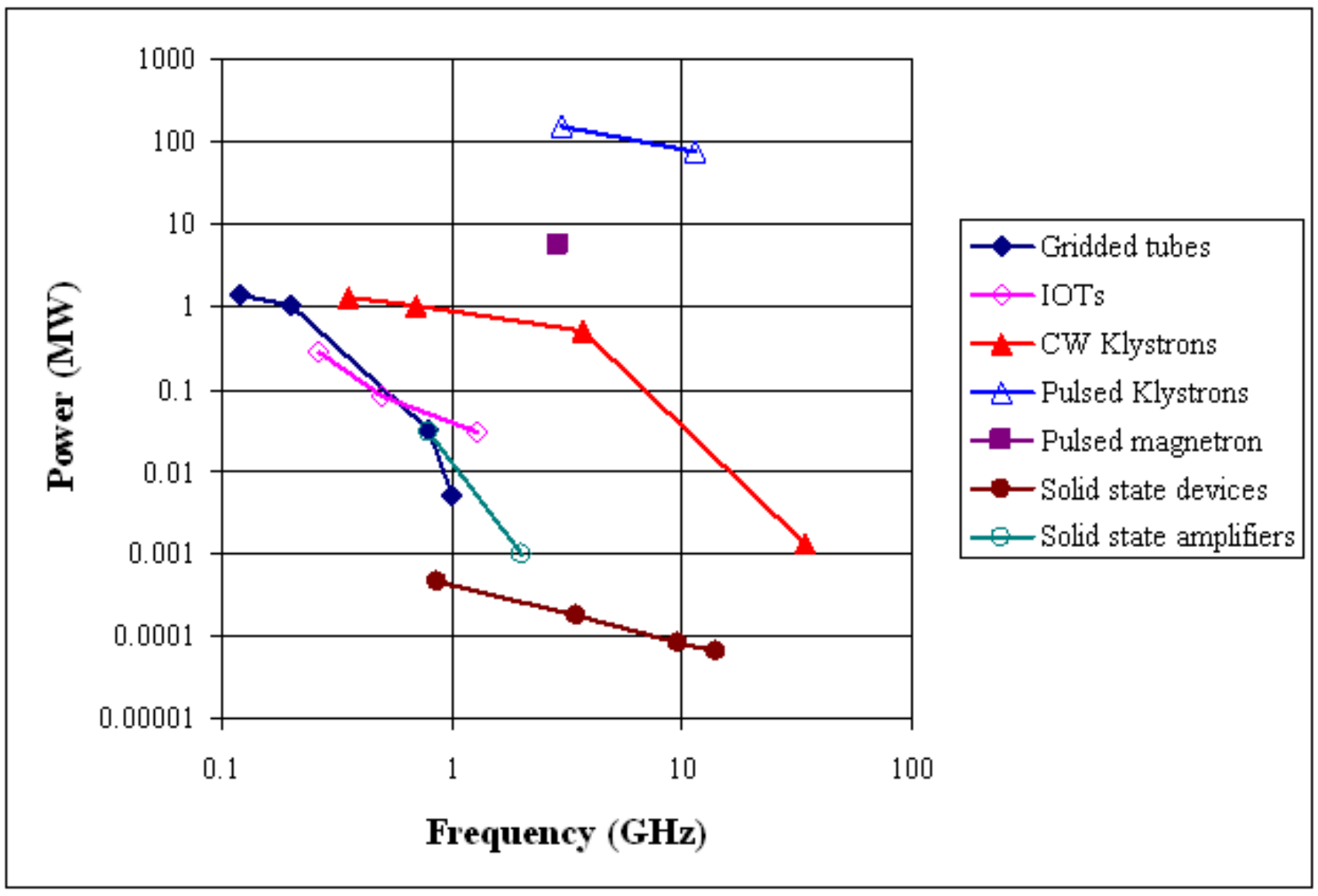

**Fig. 32:** State of the art of high-power r.f. sources